\documentclass{aa}  

\usepackage{graphicx}
\usepackage{txfonts}
\usepackage{natbib} 
\usepackage{float}
\usepackage{rotating}
\usepackage{amsmath}

\begin{document}

\title{The ALMA view of MP Mus (PDS 66): a protoplanetary disk with no visible gaps down to 4 au scales}

%
%   \subtitle{MP Mus things}

\author{ Á. Ribas \inst{1, 2} \and
  E. Macías \inst{3} \and
  P. Weber \inst{4,5,6} \and
  S. Pérez \inst{4,5,6} \and
  N. Cuello \inst{7} \and
  R. Dong \inst{8} \and
  A. Aguayo \inst{9, 10} \and
  C. Cáceres \inst{11,10} \and
  J. Carpenter \inst{12} \and
  W. R. F. Dent \inst{12} \and
  I. de Gregorio-Monsalvo \inst{2} \and
  G. Duchêne \inst{13,7} \and
  C. C. Espaillat \inst{14} \and
  P. Riviere-Marichalar \inst{15} \and
  M. Villenave \inst{16}}

\institute{Institute of Astronomy, University of Cambridge, Madingley Road, Cambridge, CB3 0HA, UK\\
    \email{ar2193@cam.ac.uk}
  \and
  European Southern Observatory, 3107, Alonso de Córdova, Vitacura, Santiago, Chile
  \and
  European Southern Observatory, Karl-Schwarzschild-Str. 2, 85748 Garching bei München, Germany
  \and
  Departamento de Física, Universidad de Santiago de Chile, Av. Victor Jara 3659, Santiago, Chile
  \and
  Millennium Nucleus on Young Exoplanets and their Moons (YEMS), Chile
  \and
  Center for Interdisciplinary Research in Astrophysics and Space Exploration (CIRAS), Universidad de Santiago
  de Chile, Estación Central, Chile
  \and
  Université Grenoble Alpes, CNRS, IPAG, 38000 Grenoble, France
  \and
  Department of Physics and Astronomy, University of Victoria, Victoria, BC, V8P 5C2, Canada
  \and
  Instituto de Física y Astronomía, Facultad de Ciencias, Universidad de Valparaíso, Av. Gran Bretaña 1111,
  Valparaíso, Chile
  \and
  Núcleo Milenio de Formación Planetaria - NPF, Valparaíso, Chile
  \and
  Instituto de Astrofísica, Universidad Andres Bello, Fernandez Concha 700, Las Condes, Santiago RM, Chile
  \and
  Joint ALMA Observatory, Avenida Alonso de Córdova 3107, Vitacura, Santiago, Chile
  \and
  Astronomy Department, University of California, Berkeley, CA 94720, USA
  \and
  Department of Astronomy and Institute for Astrophysical Research, Boston University, 725 Commonwealth
  Avenue, Boston, MA 02215, USA
  \and
  Observatorio Astronómico Nacional (OAN, IGN), Calle Alfonso XII, 3. 28014 Madrid, Spain
  \and
  Jet Propulsion Laboratory, California Institute of Technology, 4800 Oak Grove Drive, Pasadena, CA 91109, USA
}

\date{Received 09 December 2022 / Accepted 17 February 2023}

% \abstract{}{}{}{}{} 
% 5 {} token are mandatory
 
\abstract
% context heading (optional)
% {} leave it empty if necessary  
{}
% aims heading (mandatory)
{We aim to characterize the protoplanetary disk around the nearby (d$\sim$100\,pc), young solar analog MP~Mus
  (PDS~66) and to reveal any signs of planets or ongoing planet formation in the system.}
% methods heading (mandatory)
{We present new ALMA observations of MP~Mus at 0.89\,mm, 1.3\,mm, and 2.2\,mm with angular resolutions of
  $\sim$ 1\arcsec, 0.05\arcsec, and 0.25\arcsec, respectively. These data probe the dust and gas in the disk
  with unprecedented detail and sensitivity.}
% results heading (mandatory)
{The disk appears smooth down to the 4\,au resolution of the 1.3\,mm observations, in contrast with most
  disks observed at comparable spatial scales. The dust disk has a radius of 60$\pm$5\,au, a dust mass of
  $0.14_{-0.06}^{+0.11}$\,$M_{\rm Jup}$, and a mm spectral index $<2$ in the inner 30\,au, suggesting
  optically thick emission from grains with high albedo in this region. Several molecular gas lines are also
  detected extending up to 130$\pm$15\,au, similar to small grains traced by scattered light observations. Comparing
  the fluxes of different CO isotopologues with previous models yields a gas mass of $0.1-1$\,$M_{\rm Jup}$,
  implying a gas to dust ratio of 1-10. We also measure a dynamical stellar mass of
  $M_{\rm dyn}$=1.30$\pm$0.08\,$M_\odot$ and derive an age of 7-10\,Myr.}
% conclusions heading (optional), leave it empty if necessary 
{The survival of large grains in an evolved disk without gaps/rings is surprising, and it is possible that
  existing substructures remain undetected due to optically thick emission at 1.3\,mm. Alternatively, small
  structures may still remain unresolved with the current observations. Based on simple scaling
  relations for gap-opening planets and gap widths, this lack of substructures places upper limits to the
  masses of planets in the disk as low as 2\,$M_\oplus$-0.06\,$M_{\rm Jup}$ at $r > 40$\,au. The lack of mm
  emission at radii $r > 60$\,au also suggests that the gap in scattered light between 30-80\,au is
  likely not a gap in the disk density, but a shadow cast by a puffed-up inner disk.}

\keywords{Accretion disks -- Protoplanetary disks -- Planets and satellites: formation -- Stars: individual:
  MP~Mus -- Stars: pre-main sequence -- Techniques: interferometric}

\titlerunning{The ALMA view of the protoplanetary disk around MP~Mus}
\maketitle
%
%________________________________________________________________

\section{Introduction}\label{sec:intro}

Our theories of planet formation are largely informed by observations of protoplanetary disks in young, nearby
star-forming regions. Both surveys and studies of individual systems have built a general understanding of
properties such as the disk typical lifetimes, accretion rates, masses and sizes \citep[e.g., see][and other
Protostars and Planets VII chapters for a recent review of the field]{Manara2022,Miotello2022,Pascucci2022},
all of which are crucial to characterize the timescales and environment in which planets form. In recent
years, SPHERE and ALMA observations have also revealed gaps, rings and other substructures to be very common
in protoplanetary disks \citep[e.g.][]{Long2018,Andrews2018_DSHARP,Avenhaus2018}, providing new clues
about planet-disk interactions and the underlying population of newborn planets.

Although a large portion of our knowledge of protoplanetary disk properties comes from statistical analysis of
large samples, there are a few individual sources that have had a particularly high impact in our
understanding of planet formation. Perhaps the most iconic example of such a system is TW~Hya, which hosts
what is arguably the most and best studied protoplanetary disk to date. A combination of different factors
make TW~Hya a unique cornerstone in the study of planet formation. At a distance of only 60\,pc
\citep{GaiaDR3}, it is significantly closer than the nearest (140-400\,pc) star-forming regions such as
Taurus, Ophiuchus, Lupus, Chamaeleon, Upper~Scorpius or the Orion Molecular Cloud. Its proximity and almost
face-on orientation allow for very detailed studies of the disk structure: high angular resolution
observations of the gas and dust components have revealed, among other features, a concentric system of rings
and gaps \citep[including an inner gap as small as 1\,au,][]{Andrews2016}, a clump of dust at $\sim$50\,au
which may be associated with circumplanetary material \citep{Tsukagoshi2019}, a spiral structure in its gas
component \citep[][]{Teague2019}, and shadows in scattered light moving azimuthally across the disk surface,
probably cast by the inner disk \citep[][]{Debes2017}. It is also one of the few protoplanetary disks for
which a detection of hydrogen deuteride (HD) is available, allowing for a CO-independent estimate of its mass
\citep{Bergin2013} and dust-to-gas mass ratio \citep{Macias2021}, as well as the only disk in which line
polarization has been measured \citep{Teague2021}. Its 0.6\,$M_\odot$ stellar mass also makes it a great
target to better understand the early stages of the Solar System. TW~Hya greatly exemplifies the potential of
nearby protoplanetary disks for planet formation studies.

Within 100\,pc, the only other gas-rich disk around a single star is MP~Muscae (MP~Mus, PDS~66), a K1V star
\citep{Mamajek2002} located at 97.9$\pm$0.1\,pc \citep{GaiaDR3}.
It was originally identified as a classical T Tauri star by \citet{Gregorio-Hetem1992} and first believed to
belong to the $\sim$17\,Myr old Lower Centaurus-Crux association \citep[][]{Mamajek2002}, but later studies of
its kinematic properties and parallax showed it to be a member of the younger, 3-5\,Myr old $\epsilon$
Chamaeleon ($\epsilon$ Cha) association \citep{Torres2008, Murphy2013, Dickson-Vandervelde2021}. The source is
still accreting weakly \citep{Pascucci2007,Ingleby2013} at its estimated age of 7-10\,Myr, and hosts a
gas-rich disk extending up to 130\,au \citep{Kastner2010}. The SED and mid-IR spectra of the system have also
been studied \citep{Schutz2005,Bouwman2008,Cortes2009}, revealing signs of grain growth in the disk. More
recently, \citet{Wolff2016} and \citet{Avenhaus2018} presented scattered light observations from the Gemini
Planet Imager (GPI) and SPHERE/VLT, which revealed a drop in the disk brightness between 60-80\,au. If this
drop corresponds to a gap in the disk surface density, then it could be produced by the gravitational
influence of one or multiple planets, representing an excellent source to study recently formed planets in a nearby
system. With a stellar mass of 1.3\,$M_\odot$, MP~Mus may be the nearest analog to the young Solar System.

Despite the obvious interest of MP~Mus and many of its aspects being already well characterized, it still
remains comparatively unexplored at millimeter wavelengths. Here we present new observations of the system
with the Atacama Large Millimeter/submillimeter Array (ALMA), including 0.89\,mm (Band~7), 1.3\,mm (Band~6),
and 2.2\,mm (Band~4) continuum emission as well as several molecular gas lines. These observations, with an
angular resolution down to 4\,au at 1.3\,mm, provide a wealth of new information and an unprecedented view of
the system. We describe the ALMA observations as well as re-processing of ancillary SPHERE data in
Sect.~\ref{sec:observations}. We then present the results and analysis in Sect.~\ref{sec:results}, and
discuss their implications in Sect.~\ref{sec:discussion}. Finally, the main findings are summarized in
Sect.~\ref{sec:summary}.

\section{Observations and data processing}\label{sec:observations}

\subsection{ALMA observations}

MP~Mus was observed during ALMA Cycle 5 by three different programs at 0.89\,mm (Band 7), 1.3\,mm (Band 6),
and 2.2\,mm (Band 4). Project 2017.1.01687.S (P.I.: Álvaro Ribas) included observations in Bands 4 and 7,
while both projects 2017.1.01167.S (P.I.: Sebastián Pérez, part of the Disks ARound TTauri Stars with ALMA
(DARTTS-A) programme) and 2017.1.01419.S (P.I.: Claudio Cáceres) used Band 6. Observations with two antenna
configurations exist in all cases except for the Band 7 data, for which only observations with a compact
configuration are available. Also, two different executions of the Band 4 compact configuration were made. A
summary of the different datasets used and the corresponding correlator configurations is available in
Tables~\ref{tab:observations} and \ref{tab:observations_setup}. We used the standard pipeline calibration
provided by ALMA staff using {\tt CASA} \citep{CASA} version 5.1.1-5, including water vapor radiometer and
system temperature correction, as well as bandpass, amplitude, and phase calibrations. Some additional
flagging was applied to the Band 6 and Band 7 data.

\begin{table*}
\caption{Summary of ALMA observations}\label{tab:observations}
\centering
\begin{tabular}{c c c c c c c c}
  \hline\hline
  ALMA Project Code & Band & Conf. \& Baselines & Date & N$_{\rm ant}$ & Time On-Source & PWV & Flux Calibrator\\
                    & & (m) &  &  & (min) & (mm) &\\
  \hline
  2017.1.01687.S & 4 & C43-3, 15--500 & 2018 Apr 09 & 45 & 8.1 & 3.0 & J1107-4449 \\
  (P.I: \'Alvaro Ribas) & 4& C43-3, 15--500 & 2018 Apr 29 & 44 & 8.1 & 2.4 & J1427-4206 \\
                    & 4&  C43-6, 15--2500 & 2017 Dec 30   & 45 & 19.1 & 2.2 & J1617-5848 \\
                    & 7 & C43-1, 15--300 & 2018 Jul 10 & 45 & 8.8 & 0.2 & J1427-4206 \\
  \hline
  2017.1.01167.S & 6 & C43-5, 15--2400 & 2018 Jan 15 & 46 & 5.6 & 1.6 & J1427-4206 \\
  (P.I: Sebasti\'an P\'erez)   & 6& C43-8, 90--8300 & 2017 Nov 16 & 44 & 11.4 & 1.1 & J1427-4206 \\                            
  \hline
  2017.1.01419.S & 6 & C43-2, 15--300 & 2018 Jul 06 & 44 & 8.7 & 0.7 & J1427-4206 \\
  (P.I: Claudio C\'aceres) & 6 & C43-5, 15--2500 & 2017 Dec 26 & 43 & 17.1 & 0.3 & J1427-4206 \\
  \hline
  \end{tabular}
\end{table*}

\begin{table*}
\caption{Correlator configuration of the different ALMA projects used in this study}\label{tab:observations_setup}
\centering
\begin{tabular}{c c c c c c}
  \hline\hline
  ALMA Project Code & Band & Central Freq. & Bandwidth & Channels & Spectral lines\\
   & & (GHz) & (MHz) & & \\
  \hline
  2017.1.01687.S & 4 & 130.884 & 1875 & 3840 & \ldots \\
                    & & 132.801 & 1875 & 3840 & \ldots \\
                    & & 144.874 & 1875 & 3840 & DCO$^+$ (2-1), DCN (2-1), HC$_{3}$N (16-15)\\
                    & & 142.994 & 1875 & 3840 & \ldots \\  
  \cline{2-6}
                    & 7 & 330.575 & 469 & 3840 & $^{13}$CO (3-2) \\
                    & & 331.709 & 1875 & 1920 & \ldots \\
                    & & 343.292 & 1875 & 3840 & CS (7-6), HC$^{15}$N (4-3) \\
                    & & 345.783 & 469 & 3840 &  $^{12}$CO (3-2)\\
  \hline
  2017.1.01167.S  & 6 & 230.525 & 1875 & 960 &  $^{12}$CO (2-1)\\
                    & & 232.483 & 1875 & 128 & \ldots \\
                    & & 244.983 & 1875 & 128 & \ldots \\
                    & & 246.983 & 1875 & 128 & \ldots  \\
  
  \hline
  2017.1.01419.S  & 6 & 217.542 & 1875 & 128 & \ldots \\
                    & & 219.486  & 1875 & 3840 & $^{13}$CO (2-1), C$^{18}$O (2-1)\\
                    & & 230.611 & 234 & 1920 &  $^{12}$CO (2-1)\\
                    & & 231.196 & 234 & 1920 & \ldots \\
                    & & 232.791 & 1875& 128 & \ldots \\
    \hline
\end{tabular}
\end{table*}

Continuum emission from the disk was clearly detected with a high signal to noise ratio (S/N) at the three
wavelengths. Therefore, after pipeline calibration, we performed phase only self-calibration on each
individual dataset using the {\tt mtmfs} deconvolver, {\tt Briggs} weighting, a {\tt robust} value of 0.5, and
{\tt nterms}=2.  Channels with emission lines were excluded during this process.  We then re-scaled all the
data in each band to a common flux value (as reference, we chose the flux of the observation closest in time
to observations of the corresponding amplitude calibrator by the ALMA observatory), set their phase centers to
that of a Gaussian fit to the data (i.e., centered on the peak of the disk emission), and then set them to a
common coordinate to correct for pointing deviations and proper motion. In the case of the Band 6
observations, we also performed one final round of phase only self-calibration to the combined data to ensure
that they were properly aligned. The self-calibration process improved the peak S/N by factors of 2-10,
depending on the dataset.

\subsection{SPHERE scattered light observations}\label{sec:sphere_observations}

MP~Mus was observed in dual-beam polarimetric imaging mode \citep[DPI,][]{deBoer2020,vanHolstein2020} with the
InfraRed Dual-band Imager and Spectrograph (IRDIS) at SPHERE within the DARTTS-program.  The data were taken
in $J$-band and $H$-band and presented in \citet[][see this reference for details on the observational
setup]{Avenhaus2018}.  We re-reduced the DARTTS scattered light data with the reduction pipeline
IRDAP\footnote{irdap.readthedocs.io} \citep[IRDIS Data reduction for Accurate Polarimetry, version
1.3.3,][]{vanHolstein2020}, which uses a data-independent polarization model specific for the optical
instrument to correct for instrumental polarization and crosstalk.  The double-sum/double-difference technique
provides the total intensity $I$ and the linear polarization components $Q$ and $U$ (rotated by 45 \degr with
respect to each other) as data products. The total polarized intensity can be calculated from those
components:
\begin{equation}\label{equ:PI}
    PI = \sqrt{Q^2+U^2}\,.
\end{equation}
However, in the case of a single central light source and single-scattering events, it is convenient to
transform the polarized components to polar coordinates \citep[][]{Schmid2006,deBoer2020}:
\begin{equation}
 \begin{cases}
    Q_{\phi}=-Q\cos{\left(2\phi\right)}-U\sin{\left(2\phi\right)} \\ 
    U_{\phi}=+Q\sin{\left(2\phi\right)}-U\cos{\left(2\phi\right)}
 \end{cases}
\end{equation}
Here, positive $Q_\phi$ is the polarization component perpendicular to the direction of the star. Positive
$Q_\phi$ is expected to capture all stellar light that was polarized in single-scattering events with the
additional benefit of a lower noise level than $PI$ due to the lack of squared operations. On the other hand,
significant signal in $U_\phi$ or negative $Q_\phi$ can indicate regions where light is scattered more than
once \citep{Canovas2015} or where other, off-centered light sources contribute significantly to the scattering
\citep{Weber2023}.

The reprocessing of the SPHERE observations does not differ significantly from the findings of
\citet{Avenhaus2018}, but it provides additional information about the angle and degree of linear polarization
of the stellar halo. These results are discussed in Sec.~\ref{sec:sphere_results}.

\section{Results}\label{sec:results}

\begin{figure*}
  \centering
  \includegraphics[width=\hsize]{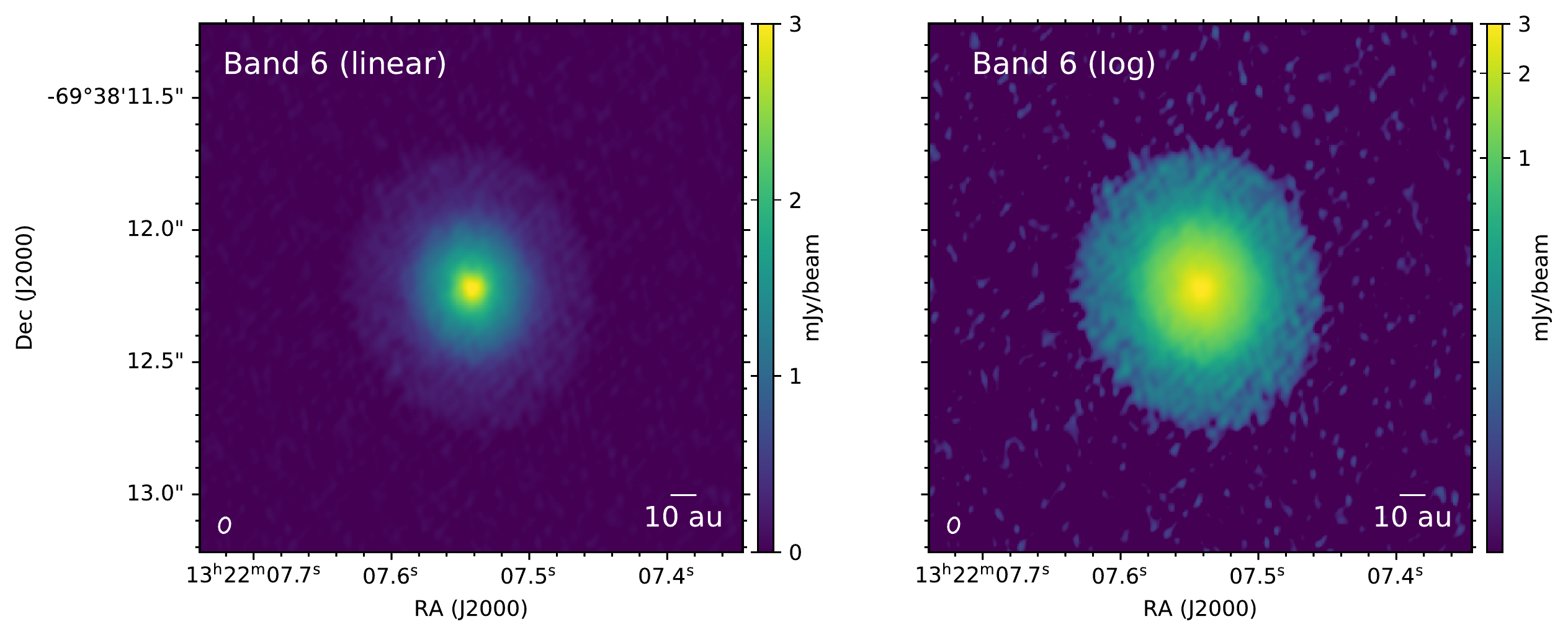}\hfill
  \caption{ALMA images of MP Mus at 1.3\,mm. These are displayed using both linear (left) and logarithmic
    (right) scales to emphasize details at different brightness levels. The 0.06\arcsec x0.04\arcsec beam is
    shown at the bottom left corners as white ellipses.}\label{fig:continuum_band6}
\end{figure*}

The new ALMA observations can be used to derive several parameters of the MP~Mus system, including dust, gas,
and stellar masses, its age, the spectral index of the millimeter continuum emission, and the overall
morphology of the disk. The SPHERE observations provide additional information about the polarization
  level and the distribution of small grains in the system. The corresponding analysis is described
throughout this section, and we provide a summary of the derived properties in Table~\ref{tab:results}.

% Results
\begin{table}
\caption{Summary of results derived for MP~Mus.}\label{tab:results}
\centering
\begin{tabular}{l l}
  \hline\hline
  Parameter & Value \\
  \hline
  $M_{*,\rm dyn}$ & $1.30 \pm 0.08 M_\odot$\\
  $L_{*}$ & $1.2\pm 0.1 L_\odot$\\
  Age\tablefootmark{\dag} & 7 - 10 Myr\\
  $V_{\rm LSR}$ & $3.98 \pm 0.04$\,km/s\\  
  $M_{\rm dust}$ & $0.14_{-0.06}^{+0.11}$\,$M_{\rm Jup}$\\
  $M_{\rm gas}$ & 0.1-1\,$M_{\rm Jup}$\\
  $R_{\rm dust}$\tablefootmark{\dag \dag} & $60 \pm 5$ au\\
  $R_{\rm gas\,(^{12}CO)}$\tablefootmark{\dag \dag} & $130 \pm 15$ au\\
  $i_{\rm dust}$ & $32 \pm 1$\degr\\
  PA$_{\rm dust}$ & $10 \pm 1$\degr\\
  $DoLP$ & $0.46 \pm 0.08$\%\\
  $AoLP$ & $98 \pm 8$\degr\\  
  \hline
\end{tabular}
\tablefoot{
  \tablefoottext{\dag}{Based on the derived luminosity and the MIST isochrones \citep{Dotter2016,
      Choi2016}.} \tablefoottext{\dag\dag}{The uncertainties correspond to beam size of the observations.}
  }
\end{table}

\subsection{Dust continuum}\label{sec:continuum}

\subsubsection{Continuum images and fluxes}\label{sec:continuum_images_fluxes}

We synthesized continuum images at 0.89\,mm, 1.3\,mm, and 2.2\,mm from the self-calibrated data described in
Sect.~\ref{sec:observations} using the {\tt tclean} algorithm with the {\tt mtmfs} deconvolver and {\tt
  nterms}=2. For each dataset at 1.3 and 2.2\,mm, the extended and compact configurations were combined to
produce the continuum images (in the particular case of the 1.3\,mm data from project 2017.1.01419.S, the
compact configuration was excluded since it was noisier and did not improve the image sensitivity).  A {\tt
  robust} value of 0.5 was used to synthesize the images to measure fluxes, resulting in beam sizes of
1.0\arcsec$\times$0.82\arcsec, 0.12\arcsec$\times$0.10\arcsec, and 0.39\arcsec$\times$0.31\arcsec at 0.89\,mm,
1.3\,mm, and 2.2\,mm, respectively.  The disk around MP Mus is clearly detected at all wavelengths (peak S/N
values of several hundreds/thousands) and is well resolved in both the 1.3\,mm and 2.2\,mm observations.
We used aperture photometry to estimate continuum fluxes from these images, obtaining 370$\pm$40\,mJy at
0.89\,mm, 148$\pm$7\,mJy at 1.3\,mm, and 49$\pm$2\,mJy at 2.2\,mm (see
Table~\ref{tab:continuum_fluxes}). These uncertainties are largely dominated by absolute calibration and not
the noise in the images. The emission at 1.3\,mm and 2.2\,mm extends up to 60\,au (0.6\arcsec), as
  determined from the 3-5\,$\sigma$ contours. To ease the comparison with other studies, we also list the
  radii enclosing 68\,\% and 90\,\% of the total flux in Table~\ref{tab:radii}. Based on the most
compact antenna configuration in each band (Table~\ref{tab:observations}), the maximum recoverable scales are
$\sim$10.8\arcsec, 2.9\arcsec, and 8.3\arcsec at 2.2\,mm, 1.3\,mm and 0.89\,mm, respectively\footnote{See ALMA
  Technical Handbook}, which are significantly larger than the observed disk size. Therefore, the observations
are likely recovering all the emission from the disk. We also produced images with a lower {\tt robust} value
of -0.5 to try to reveal small substructures, reaching angular resolutions of 0.89$\times$0.66\arcsec,
0.06\arcsec$\times$0.04\arcsec, and 0.25\arcsec$\times$0.19\arcsec at 0.89\,mm, 1.3\,mm, and 2.2\,mm
(corresponding to $\sim$75, 4, and 20\,au at 98\,pc). Interestingly, the disk appears smooth even at such
resolutions, with no clear rings, gaps, or asymmetries.  The continuum image at 1.3\,mm is shown in
Fig.~\ref{fig:continuum_band6}, and the 0.89\,mm and 2.2\,mm observations can be found in
Appendix~\ref{appendix:continuum_bands7and4}.

\begin{table*}
\caption{ALMA Continuum Fluxes}\label{tab:continuum_fluxes}
\begin{center}
\begin{tabular}{c c c c c c}
  \hline\hline
  Wavelength & Frequency & Flux & RMS & Peak S/N & Beam\\
  (mm) & (GHz) & (mJy) & ($\mu$Jy/beam) & & \\
  \hline
  0.89 & 338.187 & 370$\pm$40 & 130 & 2230 & 1.00\arcsec$\times$0.82\arcsec, PA=46\degr \\
  1.29 & 232.269 & 148$\pm$7 & 19 & 720 &  0.12\arcsec$\times$0.10\arcsec, PA=-5\degr \\
  2.17 & 137.883 & 49$\pm$3 & 14 & 1660 & 0.39\arcsec$\times$0.31\arcsec, PA=-42\degr \\
   \hline
\end{tabular}
\end{center}
\tablefoot{All fluxes estimated from images with a {\tt robust} parameter of 0.5. Uncertainties are dominated
  by the ALMA absolute flux calibration uncertainty (10\,\% for Band 7 and 5\,\% for Bands 6 and 4).}
\end{table*}

\begin{table}
\caption{Disk radius encompassing 68\,\% and 90\,\% of the total flux.}\label{tab:radii}
\begin{center}
\begin{tabular}{c c c}
  \hline\hline
  Component & $R_{\rm 68\,\%}$ (au) & $R_{\rm 90\,\%}$  (au) \\
  \hline
  Continuum (1.3\,mm) & 30$\pm$5 & 45$\pm$5 \\
  Continuum (2.2\,mm) & 30$\pm$20 & 45$\pm$20 \\
  Gas ($^{12}$CO (2-1)) &  80$\pm$15 & 110$\pm$15 \\
   \hline
\end{tabular}
\end{center}
\tablefoot{The uncertainties correspond to the beam size of each observation.}
\end{table}

\subsubsection{Disk dust mass}\label{sec:dust_mass}

Assuming that the (sub)mm emission from the disk is optically thin and isothermal, the measured flux is
linearly related to the dust mass \citep[e.g.][]{Beckwith1990}:
\begin{equation}
M_{\rm dust} = \frac{F_\nu \, d^2}{\kappa_\nu  \, B_\nu(T_{\rm dust})},
\end{equation}

where $M_{\rm dust}$ is disk dust mass, $F_\nu$ is the flux at the observed frequency $\nu$, $d$ is the
distance to the source, $\kappa_\nu$ is the dust opacity at the frequency $\nu$, and $B_\nu(T_{\rm dust})$ is
the blackbody emission at the corresponding frequency and dust temperature $T_{\rm dust}$. Since we have
observations at three different frequencies, we can compute three dust mass values. We adopted standard values
for the opacity and dust temperatures of $\kappa_{\rm 230\,GHz}$=2.3 cm$^2$/g and $T_{\rm dust}=20$\,K
\citep[e.g.,][]{Andrews2005}, and a distance of $d$=98\,pc \citep{GaiaDR3}. For the observations at 0.89\,mm
and 2.2\,mm, we computed the corresponding $\kappa_\nu$ value using a power-law dependence of the opacity with
frequency, i.e. $\kappa_\nu$ = $\kappa_{\rm 230\,GHz} \times (\nu/{\rm 230\,GHz})^{\beta}$, where $\beta$ is
between 0.0-0.6 for most protoplanetary disks \citep[e.g.,][also in agreement with the $\beta$ range of
0.1-0.4 derived for MP~Mus in Sec.~\ref{sec:grain_growth}]{Tazzari2021_lupus}. We bootstrapped the dust masses
and their uncertainties by adopting uncertainties of 5\,K for $T_{\rm dust}$ and 20\,\% for
$\kappa_{\rm 230\,GHz}$, and a uniform distribution of $\beta$ values between 0 and 0.6. The derived disk dust
masses are 0.16$_{-0.05}^{+0.1}$\,$M_{\rm Jup}$, 0.13$_{-0.04}^{+0.07}$\,$M_{\rm Jup}$, and
0.13$_{-0.04}^{+0.07}$\,$M_{\rm Jup}$
at 0.89\,mm, 1.3\,mm, and 2.2\,mm (the reported values correspond to the median and the 16\,\%, and 84\,\%
percentiles). These values are all compatible with each other, and we adopt a final dust mass value of
$M_{\rm dust}=0.14_{-0.06}^{+0.11}$\,$M_{\rm Jup}$ as the average of the three measurements.

\subsubsection{1.3\,mm continuum radial profile}\label{sec:continuum_radial}

\begin{figure*}
  \centering
  \includegraphics[width=\hsize]{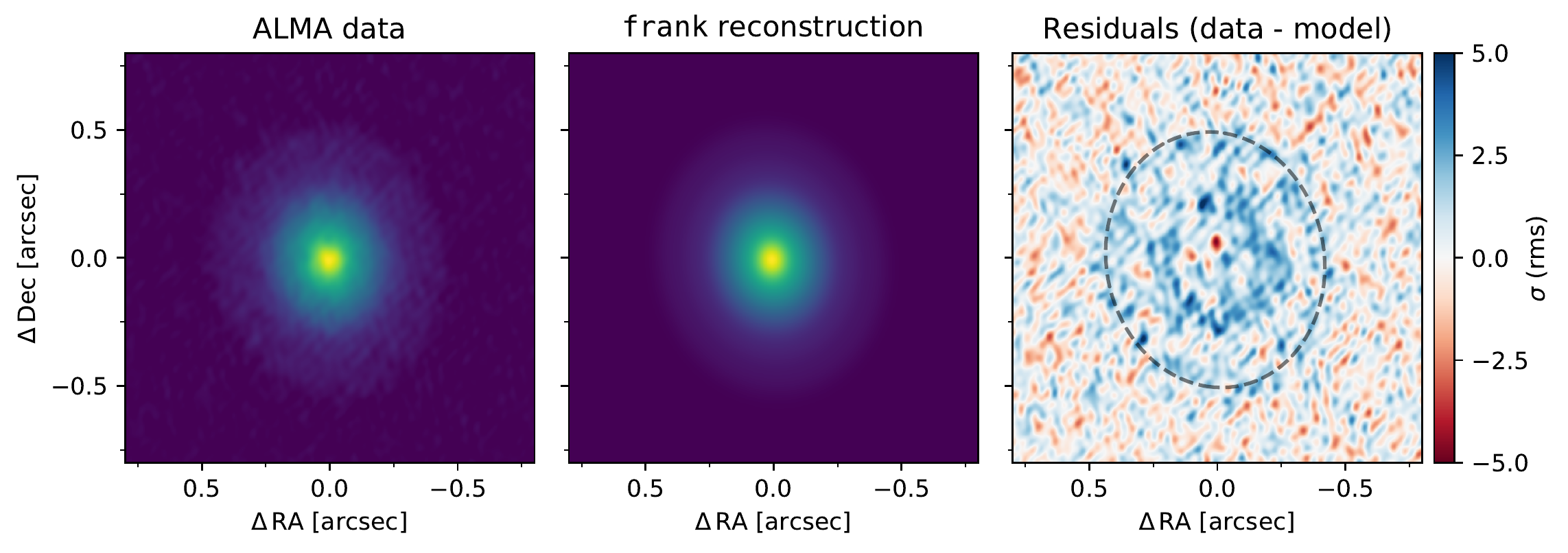}
  \caption{ALMA 1.3\,mm observations and model of MP~Mus. The observed continuum emission (left) and the resulting image
    reconstructed from the {\tt frank} radial profile (middle) are shown. The residuals (right) are displayed
    in units of the image RMS and are below the 5-$\sigma$ level.}\label{fig:frank_residuals}
\end{figure*}

\begin{figure}
  \centering
  \includegraphics[width=\hsize]{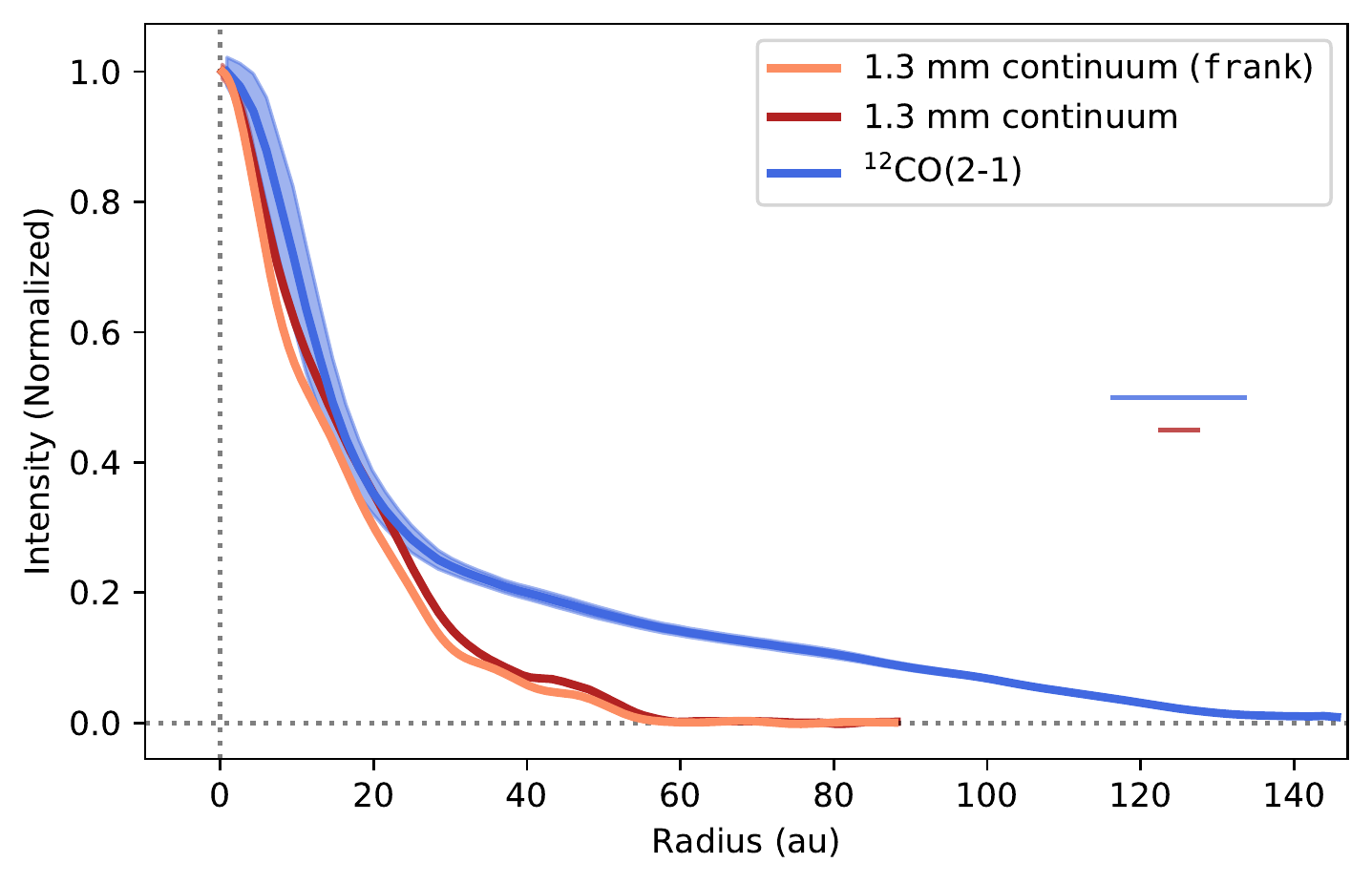}\hfill
  \caption{Radial profiles of the 1.3 mm continuum and $^{12}$CO (2-1) emission of MP~Mus. The continuum
    profile derived from the synthesized ALMA image is shown as a red line, and the orange line corresponds to
    the resulting profile from {\tt frank} \citep{frank}. The profile for the $^{12}$CO (2-1) line is also
    shown as a blue line. The 1-$\sigma$ uncertainties for the continuum and $^{12}$CO (2-1) are also plotted
    as red and blue shaded areas.  In all cases, we adopted a disk inclination and PA of 32\degr and
    10\degr. For comparison, the FWHM of the continuum and $^{12}$CO (2-1) beams are also shown as solid
    horizontal red and blue lines, respectively.}\label{fig:continuum_and_12CO_radial_profiles}
\end{figure}

To further investigate the presence (or lack) of substructures in the disk, we focused on the 1.3\,mm
continuum data since they have the highest angular resolution ($\sim$4\,au in the case of the {\tt
  robust}=-0.5 image). We first de-projected this image adopting a disk inclination of 32\degr\, and a
position angle (PA) of 10\degr\, based on a Gaussian fit to the data, in full agreement with previous
estimates from scattered light observations \citep[e.g.,][]{Schneider2014, Wolff2016, Avenhaus2018}. The
averaged radial profile was then calculated as the median intensity within concentric annuli centered on the
source. To reveal even smaller details in the disk, we also used the {\tt frank} software \citep{frank} to
reconstruct the radial profile directly from the visibilities. {\tt frank} calculates super-resolution radial
profiles of protoplanetary disks assuming azimuthal symmetry, a condition which is met in the case of MP
Mus. We tried different combinations of {\tt frank}'s $\alpha$ and $w_{\rm smooth}$ hyperparameters and found
no major differences in the resulting radial profile, so we adopted $\alpha$=1.3 and $w_{\rm smooth}=10^{-3}$
for the analysis. A comparison of the observed visibilities and the {\tt frank} fit is shown in
Fig.~\ref{fig:frank_vis_fit}.  The inclination and PA derived from {\tt frank} are in complete agreement with
the values adopted earlier. Both the profile extracted directly from the image and the one from {\tt frank}
reveal radially decreasing emission extending up to $\sim$60\,au, with changes in the slope at $\sim$10 and
30\,au as well as a plateau between 30-40\,au, and bump in the outermost region which may suggest the presence
of a low-contrast gap and small, barely resolved ring. However, no clear signatures of substructures are found
down to a 4\,au scale.  We also produced a residual map by extending the radial profile from {\tt frank}
azimuthally, projecting the resulting image with the corresponding disk inclination and orientation, and
convolving it with the observed beam before subtracting it from the observations.  The residuals
(Fig.~\ref{fig:frank_residuals}) are all below the 5-$\sigma$ level and do not reveal any azimuthal
substructures. The resulting radial profiles are shown in Fig.~\ref{fig:continuum_and_12CO_radial_profiles}.

\subsubsection{(Sub)mm spectral indices}\label{sec:spectral_indices}

\begin{figure}
  \centering
  \includegraphics[width=\hsize]{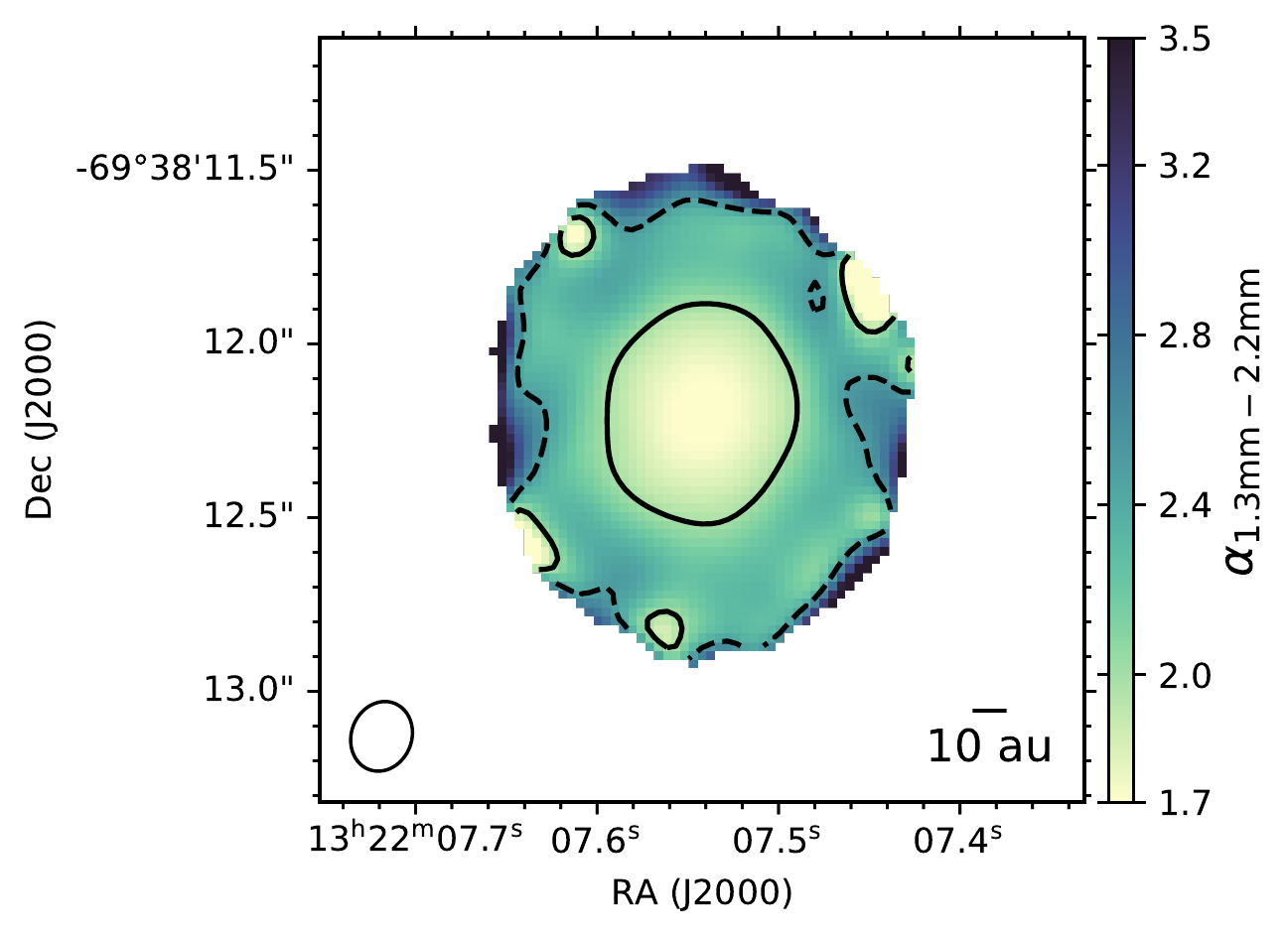}
  \includegraphics[width=\hsize]{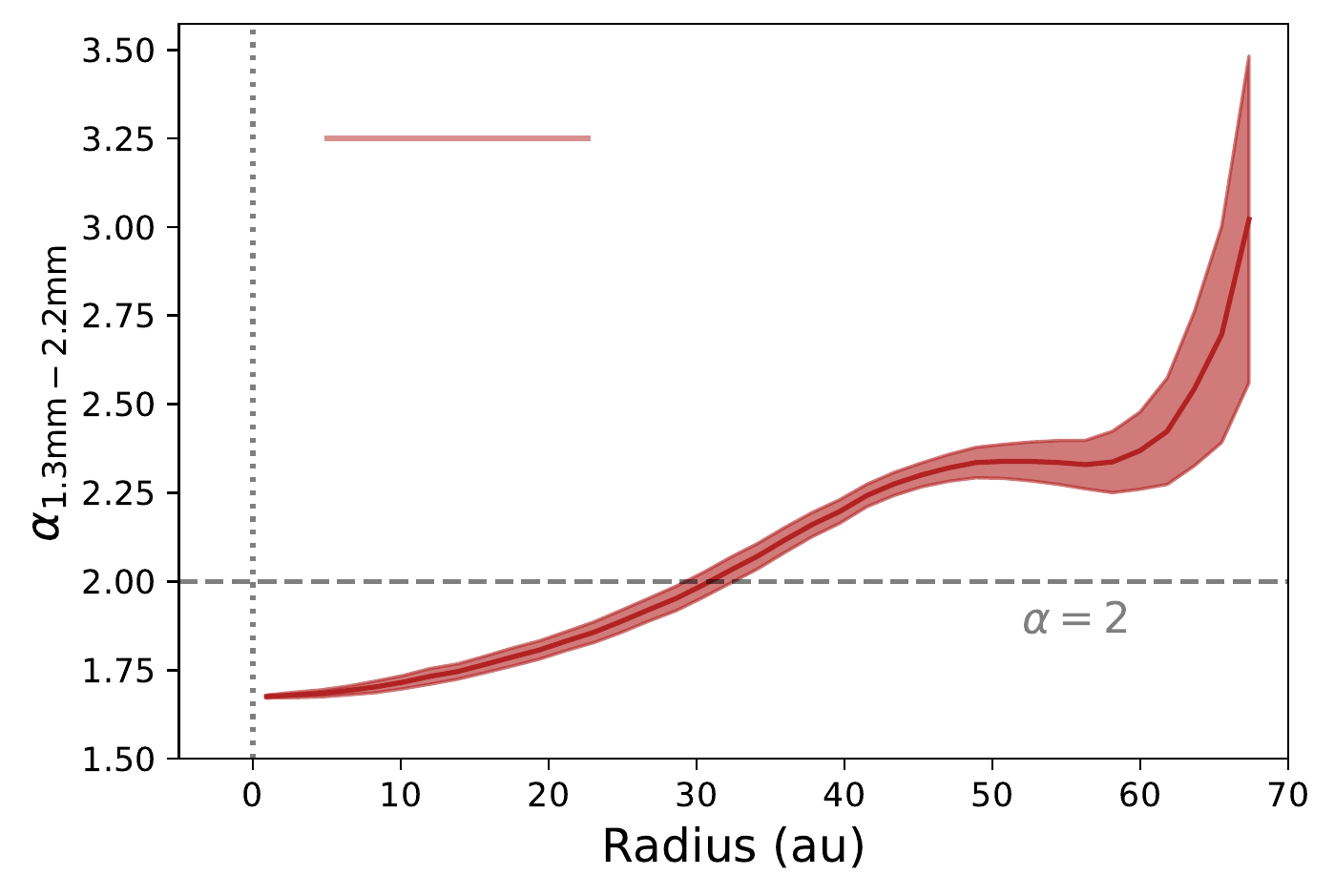}
  \caption{Derived millimeter spectral index for MP~Mus between 1.3 and 2.2\,mm. Top: Spectral index map. Only
    pixels with S/N$>$5 are considered. The solid and dashed contours correspond to
    $\alpha_{\rm 1.3 - 2.2\,mm}$ of 2 and 2.5, respectively. The image beam is shown on the bottom left
    corner. Bottom: Corresponding de-projected $\alpha_{\rm 1.3 - 2.2\,mm}$ radial profile and uncertainties
    (solid red line and area). The beam FWHM is shown as a horizontal red line. The dashed line marks the
    $\alpha=2$ transition.}\label{fig:alpha_map_and_profile}
\end{figure}

The spectral index ($\alpha_{\rm mm}$) of optically thin (sub)mm emission from protoplanetary disks depends on
the size of dust grains in them, and has been used in the past to investigate grain growth in disks in several
star-forming regions \citep[e.g.][]{Ricci2010_Taurus, Ricci2010_Ophiuchus, Ribas2017,
  Tazzari2021_lupus}. Using the derived continuum fluxes, we computed three spectral indices in different
wavelength ranges: $\alpha_{\rm 0.89 - 1.3\,mm}=2.4 \pm 0.3$, $\alpha_{\rm 1.3 - 2.2\,mm}=2.12 \pm 0.11$, and
$\alpha_{\rm 0.89 - 2.2\,mm}=2.25 \pm 0.13$ (including the absolute calibration uncertainties of ALMA). 
These values were computed from the integrated fluxes and they reflect the average spectral index in the disk
only, but the spectral index is expected to vary spatially as a result of factors such as radial changes in
the optical depth and grains sizes. To investigate such spatial variations, we combined the resolved 1.3\,mm
and 2.2\,mm observations to produce a resolved map of the spectral index. During this process, the most
extended configuration of the available Band 6 data was excluded to avoid problems with very different
coverage of the uv-plane at different bands. The 1.3\, and 2.2\,mm data were jointly imaged with the {\tt
  tclean} algorithm, the {\tt mtmfs} deconvolver, and {\tt nterms}=2, and we then used the resulting {\tt
  alpha} image as the spectral index map. After various tests we adopted a {\tt robust} parameter of 0.0 in
this case, which yielded a beam of 0.2\arcsec $\times$0.17\arcsec. The derived spectral index map and its
radial profile are shown in Fig.~\ref{fig:alpha_map_and_profile}. As expected, the value of
$\alpha_{\rm 1.3 - 2.2\,mm}$ is not constant throughout the disk and increases as a function of radius,
ranging from $\sim$1.7 in the inner regions to $\sim$3 in the outer parts of the disk. These results are
discussed in further detail in Sec.~\ref{sec:grain_growth}.

\subsection{Gas lines}\label{sec:gas}

\subsubsection{Line cubes, fluxes, and morphology}\label{sec:gas_lines}

The observations in this study cover multiple molecular gas emission lines, which were imaged using {\tt
  tclean} after applying the self-calibration solutions and re-centering derived in
Sec.~\ref{sec:observations} and subtracting the corresponding continuum. We detected $^{12}$CO~(3-2),
$^{13}$CO~(3-2), and CS~(7-6) in Band 7, $^{12}$CO~(2-1), $^{13}$CO~(2-1), and C$^{18}$O~(2-1) in Band 6, and
DCO$^+$~(2-1), DCN~(2-1), and HC$_{3}$N~(16-15) in Band 4. HC$^{15}$N (4-3) was also tentatively detected in
the Band 7 data. We used CASA to produce the zero-th and first moments for each line, and applied Keplerian
masking during the process to minimize noise from signal-free areas \citep[e.g.,][]{Salinas2017}. The moments
and spectra for the$^{12}$CO~(2-1), $^{13}$CO~(2-1), and C$^{18}$O~(2-1) are shown in Fig.~\ref{fig:lines_B6},
and similar figures for the remaining lines are provided in Appendix~\ref{appendix:lines}. The line fluxes
measured from the zero-th moments are listed in Table~\ref{tab:line_fluxes}. Note that each project used
different antenna and correlator configurations, so the spectral and spatial resolutions are different for
each line (also listed in Table~\ref{tab:line_fluxes}). In the case of the Band~6 observations, we only used
the observations from project 2017.1.01419.S to generate the cubes (data from 2017.1.01167.S have a
significantly higher angular resolution which results in a much lower sensitivity per channel, as well as a
coarser spectral resolution). We also used different weighting depending on the line as a compromise between
angular resolution and sensitivity: all lines in Band~4 as well as HC$^{15}$N~(4-3) in Band~7 were imaged with
{\tt natural} weighting to maximize the S/N, and the remaining lines were imaged using a {\tt robust} value of
0.5.

\begin{table*}
\caption{Gas Line Detections and Properties}\label{tab:line_fluxes}
\begin{center}
\begin{tabular}{l c c c c c}
  \hline\hline
  Line & Band & Rest. Frequency & Line Flux & Beam size & Spectral resolution\\
          & & (GHz) & (Jy km/s) &  (\arcsec) & (km/s)\\
  \hline
   $^{12}$CO (3-2)        & 7 & 345.796  & 10 $\pm$ 1 & 1.06$\times$0.86 & 0.11 \\
   CS (7-6)                   & 7 & 342.883 & 0.43 $\pm$ 0.05 &  1.05$\times$0.87 & 0.43 \\
   HC$^{15}$N (4-3)     & 7 & 344.200 & 0.04 $\pm$ 0.02 & 1.14$\times$0.97 & 0.85\\
   $^{13}$CO (3-2)        & 7 & 330.588 & 1.8 $\pm$ 0.2 & 1.10$\times$0.91 & 0.11\\
  $^{12}$CO (2-1)         & 6 & 230.538  & 4.6 $\pm$ 0.2 & 0.23$\times$0.20  & 0.16\\
   $^{13}$CO (2-1)        & 6 & 220.399 & 0.79 $\pm$ 0.05 & 0.24$\times$0.21  & 0.66 \\
   C$^{18}$O (2-1)        & 6 & 219.560 & 0.21 $\pm$  0.02 & 0.24$\times$0.21 & 0.67\\
   HC$_3$N (16-15)      & 4 & 145.561 & 0.34 $\pm$ 0.02 & 0.50$\times$0.40 & 1.01\\
   DCN (2-1)                & 4 & 144.828 & 0.08 $\pm$ 0.01  & 0.50$\times$0.40 & 1.01 \\
   DCO$^+$ (2-1)         & 4 & 144.078 & 0.11 $\pm$ 0.01  & 0.50$\times$0.40  & 1.02\\
   \hline
\end{tabular}
\end{center}
\tablefoot{Uncertainties include the absolute ALMA flux calibration uncertainties.}
\end{table*}

The observations show gas emission at velocities from $\sim$-5 to 13\,km/s, and we measured a systematic
velocity of 3.98 $\pm$ 0.04\,km/s (local standard of rest) using a Keplerian disk model (see
Sec.~\ref{sec:dynamical_mass}), similar to previous estimates \citep[e.g.,][]{Kastner2010}.
As shown in Figs.~\ref{fig:lines_B6}, \ref{fig:lines_B7}, and \ref{fig:lines_B4}, the emission shape of most
lines is that of a full disk., i.e. no clear gaps or rings are found. Exceptions are DCO$^+$ (2-1), which
shows a clear ring-like morphology with a gap radius of $\sim$20\,au \citep[a morphology commonly
  observed for this line, e.g.,][]{Huang2017}, and DCN (2-1) and HC$^{15}$N (4-3), where the low S/N prevents
any reliable estimate of their morphology. The gaseous disk extends up to 130\,au (1.3\arcsec) in
  $^{12}$CO (2-1) based on the 3-5\,$\sigma$ contours, in agreement with previous studies using unresolved
APEX observations of the $^{12}$CO (3-2) line \citep[when corrected from the updated Gaia
distance,][]{Kastner2010}. We also provide the radii encircling 68\,\% and 90\,\% of the total $^{12}$CO
  (2-1) emission in Table~\ref{tab:radii}. The gas radius is $\sim$twice that of the dust disk
(60\,au, Sec.~\ref{sec:continuum}) yielding a ratio of the gas and dust radii similar
to those found for disks in Lupus~\citep{Ansdell2018, Sanchis2021}. A comparison of the de-projected radial
profiles of the continuum and $^{12}$CO (2-1) emission is shown in
Fig.~\ref{fig:continuum_and_12CO_radial_profiles}. We note that none of the ALMA projects used in this work
aimed at studying the chemistry of MP~Mus and yet these observations detected various molecular emission lines
of multiple species, evidencing the large potential of this system for future astrochemical studies of
protoplanetary disks.

\begin{figure*}
  \centering
  \includegraphics[width=\hsize]{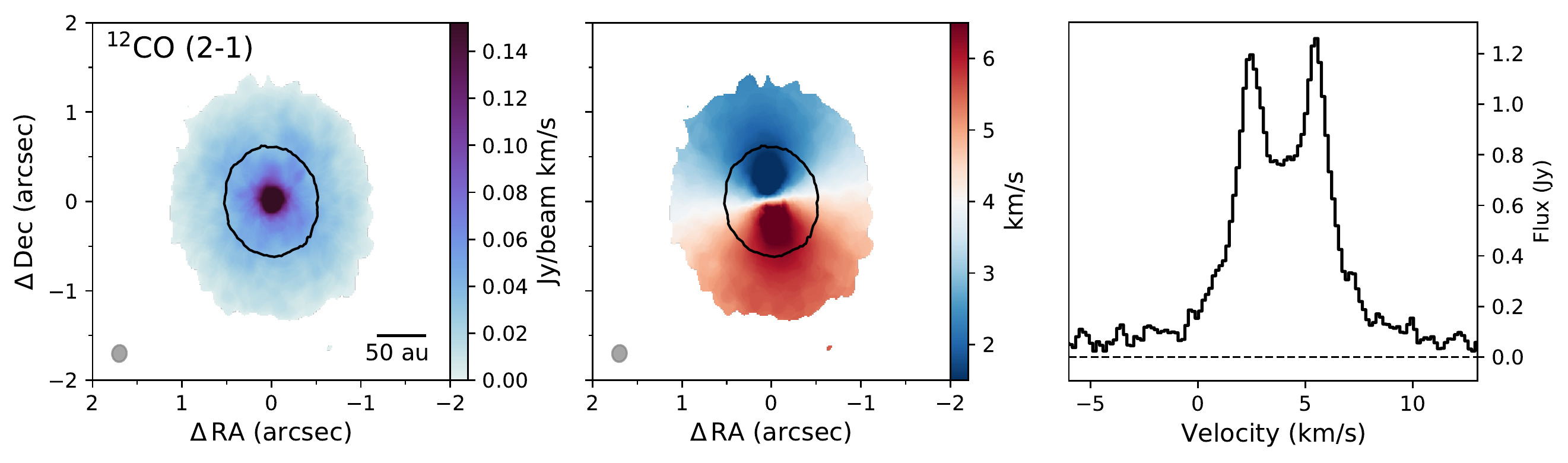}
  \includegraphics[width=\hsize]{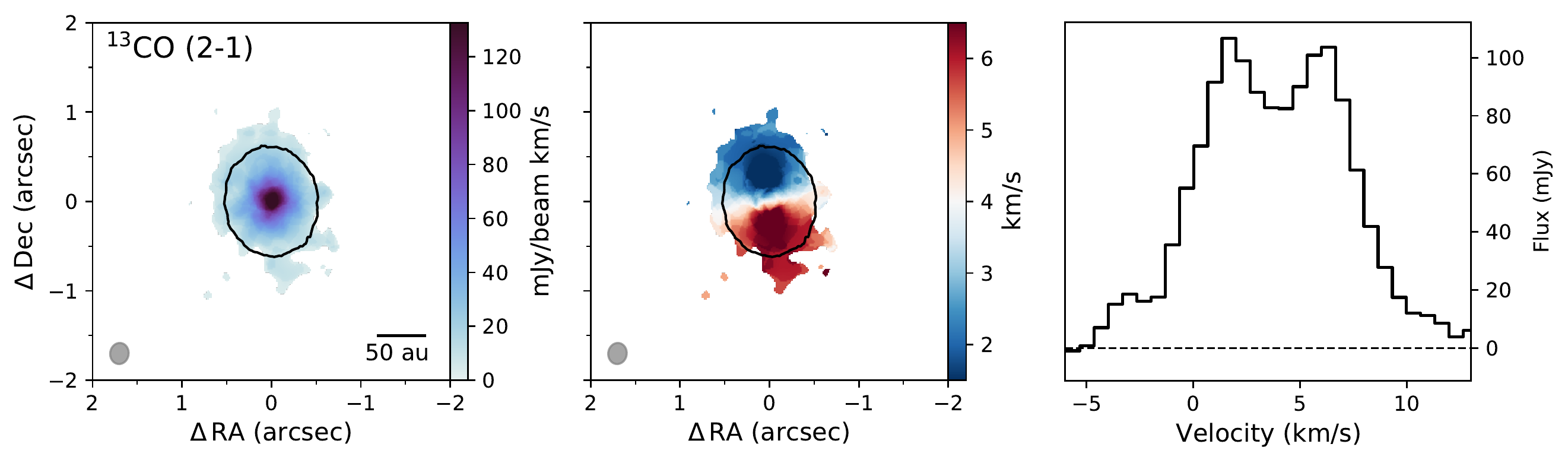}
  \includegraphics[width=\hsize]{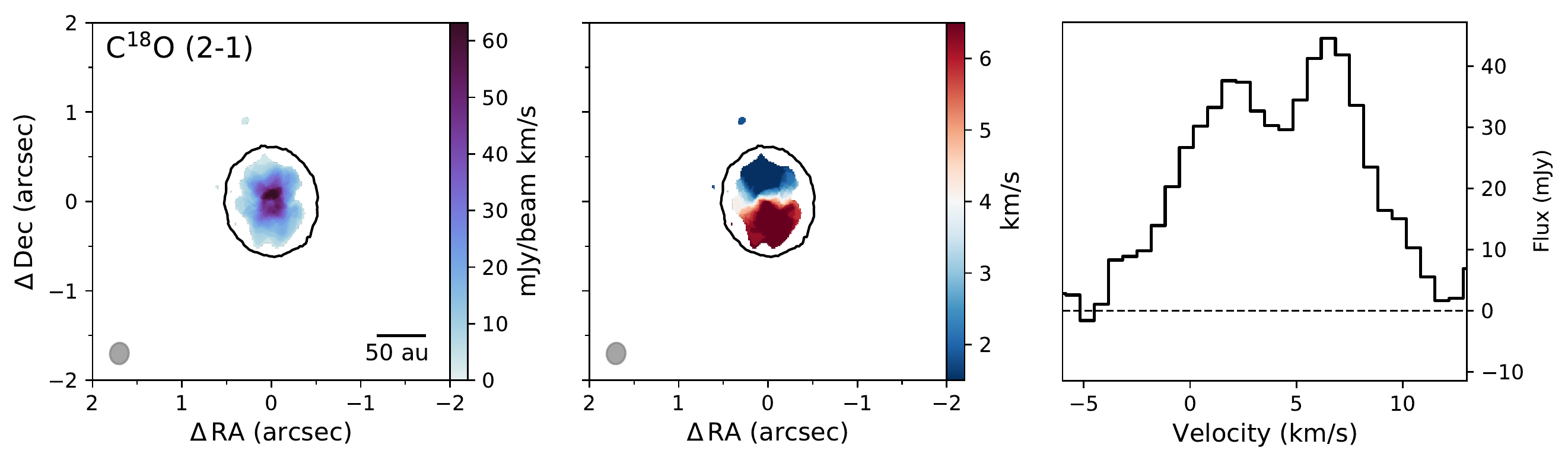}
  \caption{ALMA Band~6 observations of CO isotopologues from the protoplanetary disk around MP~Mus. The
    top, middle, and bottom rows correspond to ${^{12}}$CO (2-1), ${^{13}}$CO (2-1), and C${^{18}}$O (2-1),
    respectively. The zero-th (left column) and first (middle column) moments are shown, together with the
    5$\sigma$ contour of the 1.3\,mm continuum as a black line. The corresponding spectra are displayed in the
    right column.}\label{fig:lines_B6}
\end{figure*}

\subsubsection{Dynamical stellar mass and disk kinematics}\label{sec:dynamical_mass}

\begin{figure*}
  \centering
  \includegraphics[width=0.33\hsize]{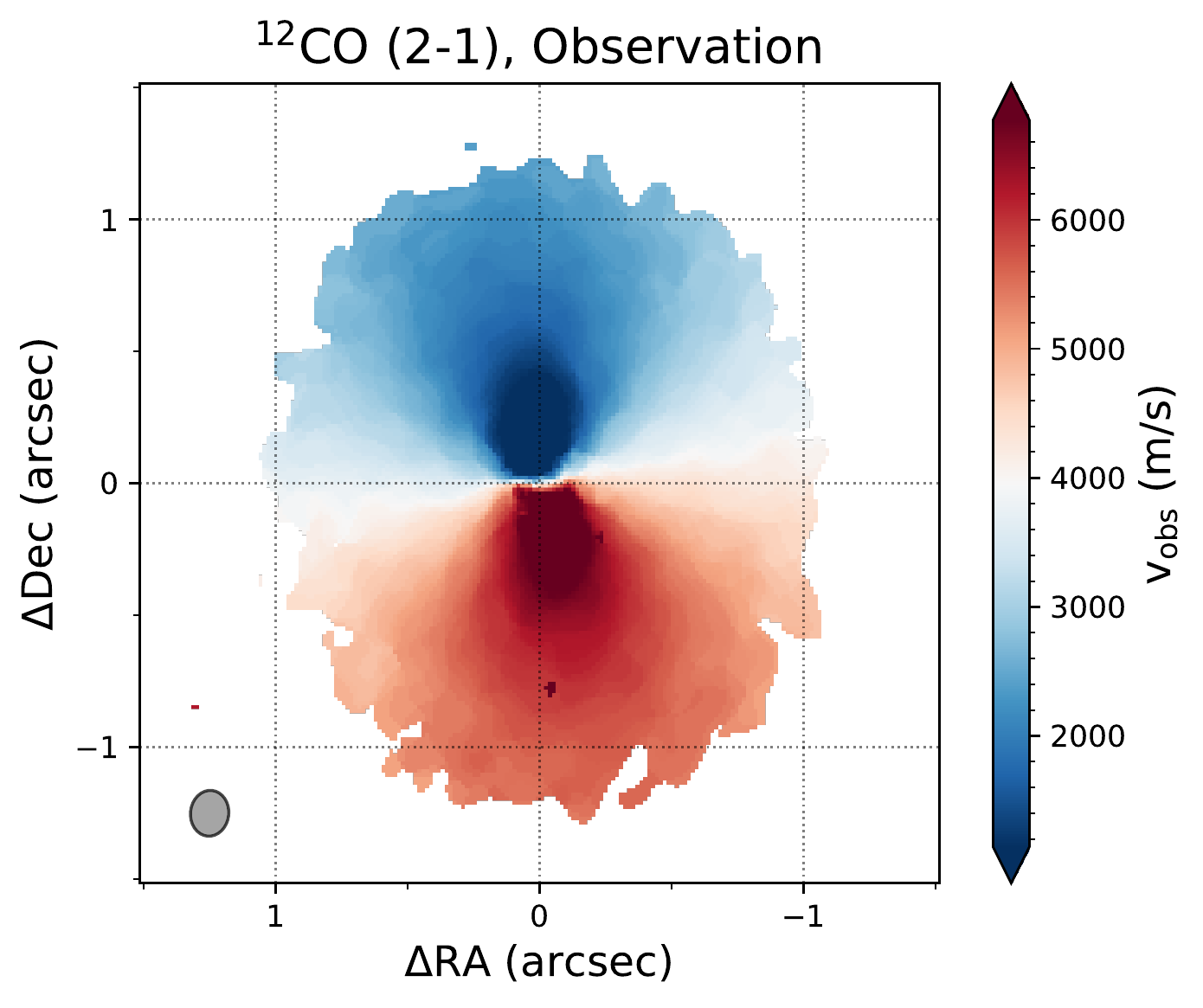}
  \includegraphics[width=0.33\hsize]{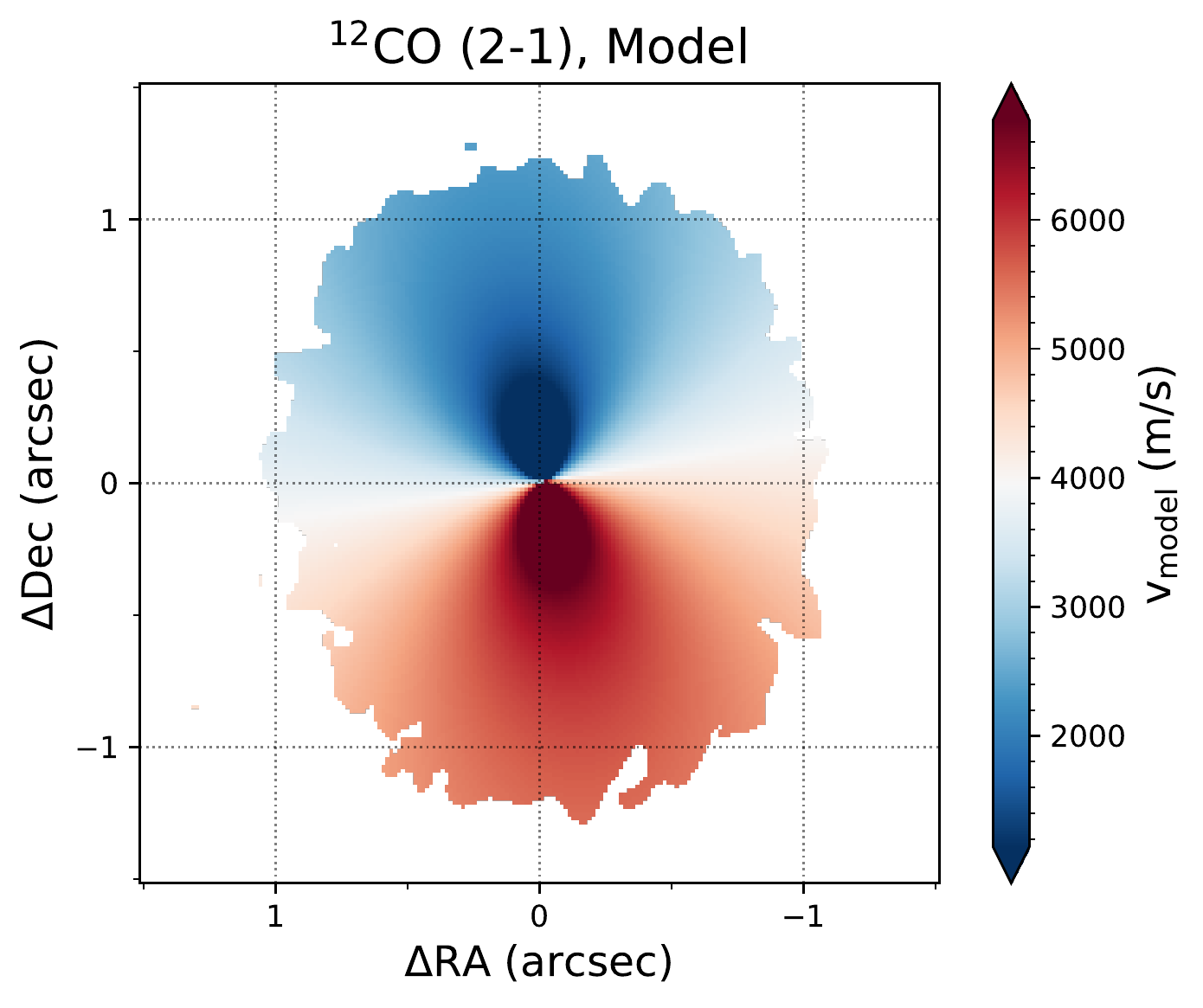}
  \includegraphics[width=0.33\hsize]{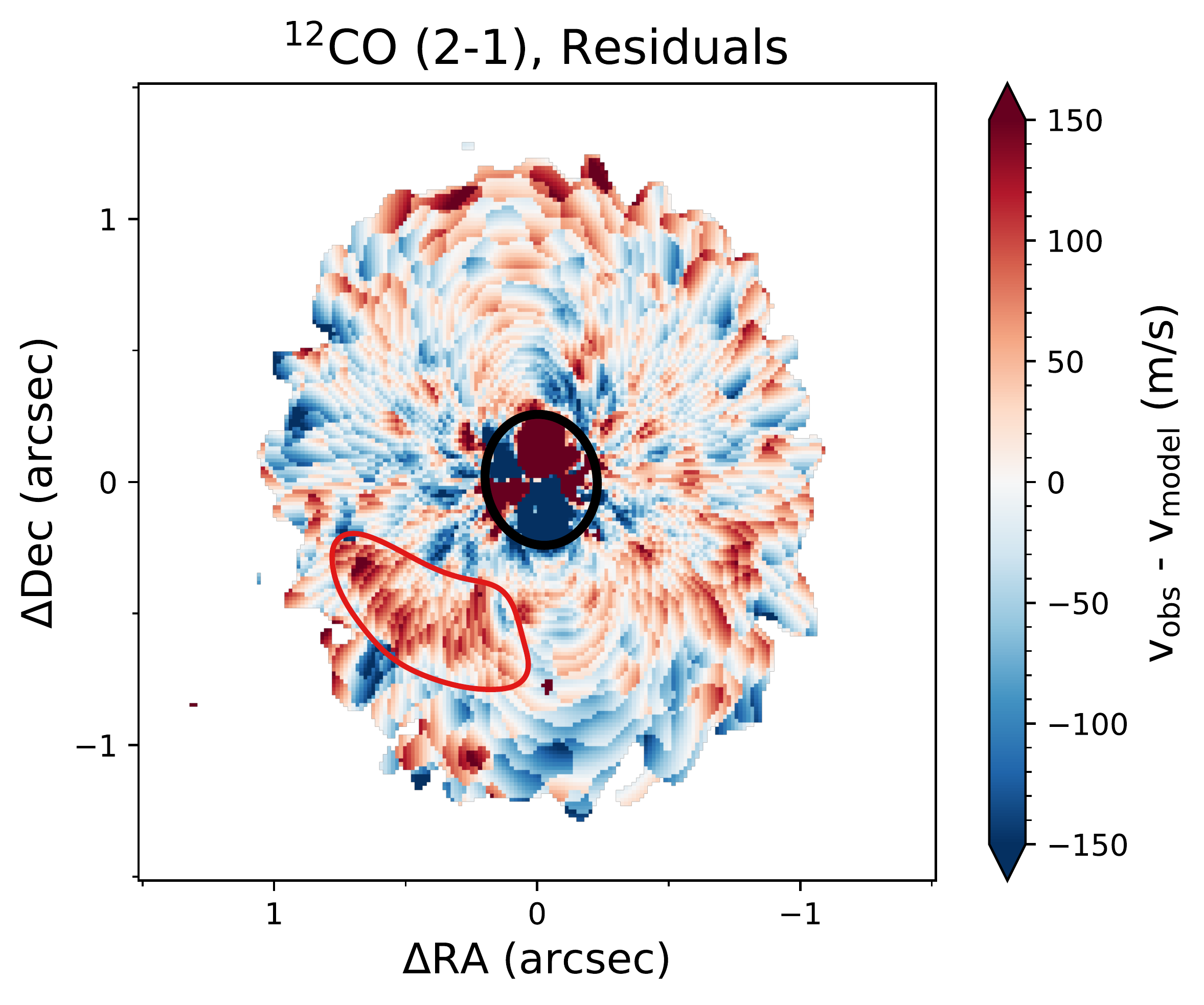}
  \caption{Keplerian fit to the ${^{12}}$CO (2-1) line emission of MP~Mus using {\tt eddy}. The observed first moment
    (left), model (middle), and corresponding residuals (right) are shown. The 0.17\arcsec x 0.15\arcsec beam
    is shown on the bottom left corner. The area inside the black ellipse shown in the residuals was
      masked during the fit. The tentative localized residuals mentioned in the text are also marked with a
      red line. Only the extended configuration from 2017.1.01419.S was used for this analysis.}\label{fig:Keplerian_fit}
\end{figure*}

Emission from gas lines experiences Doppler shifts due to the disk rotation, and can thus be used to estimate
stellar mass that independently of theoretical isochrones and stellar evolution models. For this purpose, we
used the {\tt eddy} software \citep{eddy} to model the first moment map of the $^{12}$CO (2-1) emission using
a Keplerian rotation profile. This line was chosen as it offers the best compromise between S/N and the
available spatial resolution. Only the extended configuration of 2017.1.01419.S was used for this purpose
since the compact one is significantly noisier, but tests including this second observations yielded noisier
but completely compatible results (as mentioned in Sec.~\ref{sec:gas_lines}, the observations from
2017.1.01167.S have a coarser spectral resolution and significantly less sensitivity per channel, so they were
not included in this analysis). Given the high S/N of the observations, we imaged the line with a {\tt
  robust}=0.0 weighting (0.17\arcsec x0.15\arcsec beam) to improve the angular resolution. The first moment
map and its corresponding uncertainties were then calculated using the {\tt bettermoments} software
\citep[][]{bettermoments}. In Sec.~\ref{sec:continuum_radial} we derived a disk inclination and PA of 32\degr
and 10\degr based on the dust continuum observations which have a significantly higher angular resolution, so
we kept those values fixed during the fitting \footnote{{\tt eddy} defines the PA with respect to the
  red-shifted semi-major axis, so the adopted value was 190\degr.}. We performed various tests during this
process, including the use of first moment maps computed with the {\tt CASA immoments} task instead,
down-sampling the map to the beam size to ensure that only spatially-independent pixels are fit, masking
the inner 0.3\arcsec of the disk, and modifying the disk inclination by $\pm$1\degr to account for this
  uncertainty in the final results. In all cases, we obtain very similar stellar mass values for MP~Mus, and
adopt a final value of $M_*$=1.30 $\pm$ 0.08\, $M_\odot$. This value is slightly higher than the 1.2\,
$M_\odot$ value in the literature based on pre-MS evolutionary tracks \citep[e.g.][]{Mamajek2002}, but in
complete agreement when updating the luminosity with the new Gaia distance. This process also yields the
aforementioned systemic velocity of 3.98$\pm$0.04\,m/s (local standard of rest) for MP~Mus. The $^{12}$CO
(2-1) map used, model, and residuals are shown in Fig~.\ref{fig:Keplerian_fit}.

The $^{12}$CO (2-1) data and moment maps show no obvious flaring, and the disk back surface is also not
  visible. This suggests a very flat morphology of the gaseous disk, in agreement with the results from the
  scattered light observations \citep{Avenhaus2018}. To further explore this, we also tested the elevated
  surface emission prescription from a tapered disk in {\tt eddy}, which parametrizes the emission height as a
  function of radius following $z(r) = z_0 (r/1\arcsec)^{\psi} \exp{[-(r/r_{tapper})^{q_{tapper}}]}$ (where
  $z_0$ is the disk aspect ratio at 1\,\arcsec, $\psi$ is the disk flaring, and $r_{tapper}$ and $q_{tapper}$
  account for the disk tappering). This resulted in a stellar mass similar to that of the geometrically thin
  disk case, largely unconstrained values for $\psi$, $r_{tapper}$ and $q_{tapper}$, and a $z_0$ value of
  0.1$\pm$0.1, showing that the disk is indeed significantly flat.

There are two interesting features worth noticing from the fitting of Keplerian rotation profiles. Firstly,
the $^{12}$CO (2-1) first moment shows a tentative twist in the inner disk (i.e., change in the PA in the
inner regions), and the residuals are structured and quite high in this area (up to $\sim$1\,km/s). These
residuals could indicate the presence of a warped, twisted, or misaligned inner disk
\citep[e.g.][]{Marino2015,Casassus2015, Min2017, Benisty2018, Mayama2018, Bohn2022}. Such structures can cast
azimuthal shadows on the disk that can be detected in scattered light data, and \citet{Wolff2016} tentatively
detected such a shadow in GPI observations of MP~Mus. Later observations with SPHERE did not recover this
feature, and \citet{Avenhaus2018} suggested that an imperfect correction of instrumental or interstellar
polarization may create a similar effect. Alternatively, variability in the inner disk could also change the
appearance of a shadow considering the 2 year separation between both observations. The angle of linear
polarization of the unresolved stellar signal derived from the SPHERE data (see Sec.~\ref{sec:sphere_results})
also suggests that the inner and outer disks are aligned. Secondly, we also find some localized residuals in
the first moment at radii 50-100\,au between PA 110-170\,\degr (Fig.~\ref{fig:Keplerian_fit}). A detailed
analysis of this structure is beyond the scope of this work, but its localized nature is suggestive of the
velocity perturbations attributed to unseen planets in other systems \citep[e.g.,][]{Pinte2018, Perez2018,
  Teague2018}. The emission in some other channels also displays tentative deviations from the expected Keplerian
profile that are typically interpreted as such kinks (e.g., see the $^{12}$CO (2-1) NE emission in
Fig.~\ref{fig:SPHERE_results}), but the sensitivity and resolution of our observations do not allow us to draw
any conclusion. Further observations of the $^{12}$CO (2-1) and optically thinner tracers at higher spatial
and spectral resolution are needed to better characterize the inner regions and possible deviations from
Keplerian rotation in MP~Mus.

\subsubsection{Disk gas mass and gas-to-dust ratio}\label{sec:gas_mass}

The bulk of mass in protoplanetary disks is in gaseous form. However, in contrast with the dust
mass which can be derived (or at least approximated) from mm fluxes, such estimates are much more complex
for the gas. A number of methods can be used for this purpose, but reliable gas mass measurement usually
require detailed modeling using chemical networks, radiative transfer, a good knowledge of the disk
structure, and resolved observations of multiple emission lines (together with a large number of assumptions
regarding chemical abundances). Such a study is outside the scope of this work, but we can obtain some
order-of-magnitude estimates by comparing the observed $^{13}$CO (2-1) and C$^{18}$O (2-1) line luminosities
(9.5$\times$10$^{4}$ Jy km/s pc$^{2}$ and 2.5$\times$10$^{4}$ Jy km/s pc$^{2}$, respectively) with model grids.

\citet{WilliamsBest2014} produced a suite of disk models with various properties and derived the resulting
line fluxes for different CO lines. Their modeling did not include the selective photodissociation for the
$^{13}$CO and C$^{18}$O isotopologues and, instead, they included this effect by calculating half of their
models with the usual CO abundances and the other half with a [C$^{18}$O]/[$^{13}$CO] ratio three times
lower. Comparing the observed luminosities of $^{13}$CO (2-1) and C$^{18}$O (2-1) in MP~Mus with their grid of
models (focusing on the M$_*$=1\,M$_\odot$ and inclination=10\degr models since they are the closest to this
system) results in gas masses between 3$\times$10$^{-4}$\,M$_\odot$ and 1$\times$10$^{-3}$\,M$_\odot$ for the
cases without and with C$^{18}$O depletion, respectively. Similarly, \citet{Miotello2016} investigated the
dependence between disk masses and the luminosities of various CO lines for different disk properties using a
grid of models including chemical modeling. Comparing the derived line luminosities with their grid of
models \citep[see Fig.7 in][]{Miotello2016} places the gas mass of MP~Mus between 10$^{-4}$ and
10$^{-3}$\,M$_\odot$, depending on whether isotope-selective processes are considered or not.

Although these models are not specifically tailored to MP~Mus, these general comparisons suggest that its
total gas mass is M$_{\rm disk}$=1$\times$10$^{-4} - 10^{-3}$\,M$_\odot$ (0.1-1\,M$_{\rm Jup}$). Taken at face
value and combined with the dust mass estimate in Sec.~\ref{sec:dust_mass}, this implies a global gas-to-dust
ratio of 1-10, lower than the standard value of 100 in the ISM. We note that these values are derived
  using the global dust and gas masses but, since the dust disk is considerably smaller than the gas one, this
  implies an even lower gas-to-dust ratio in the area where both gas and large dust grains are
  present. MP~Mus then joins the increasing number of sources with a gas-to-dust ratio below 100, which
recent ALMA surveys have shown to be common in protoplanetary disks: as an example, \citet{Miotello2017} found
that 23 out of 34 disks surveyed in the Lupus star-forming region showed gas-to-dust ratios below
10. Traditionally, these values are interpreted as a signpost of disk evolution, where this ratio decreases
over time as gas dissipates in the disk while dust grains remain. Such a scenario is reasonable for an evolved
and flat disk such as MP~Mus. However, another possible explanation is that the CO abundance in disks is lower
than expected and results in fainter line emission, as suggested by mass measurements based on HD for TW~Hya,
GM~Aur and DM~Tau \citep{Bergin2013,McClure2016} and in Lupus using N$_2$H$^+$
\citep{Anderson2022}. Unfortunately, testing this hypothesis requires reliable gas mass estimates that are
independent of the CO abundance, which are extremely challenging and not yet available for MP~Mus. The actual
reason for the low CO-based gas masses in disks is still an open question for planet formation theories.

\subsection{Scattered light results}\label{sec:sphere_results}
  
\citet{Avenhaus2018} presented the DPI data taken in $J$- and $H$-band within the DARTSS program.  These data
were reduced by minimizing the non-azimuthal component of the radiation field, as an accurate polarization
model for the IRDIS instrument was not available at the time.  In contrast, we employed the
instrument-specific polarization model used within the IRDAP pipeline, which allowed us to subtract
instrumental polarization and polarization crosstalk from the image without any assumption about the data.
Besides the scattered light of the disk, the image then still contains a background level of polarization and
some unresolved polarization carried within the stellar halo.  IRDAP measures the background polarization in
regions where the signal intensity is at noise level and subtracts this background from the entire image.  To
estimate the stellar polarization, we measured $Q$ and $U$ from the stellar halo in image areas seemingly free
of disk-scattered light (in an annulus mask of 1.72\arcsec - 2.08\arcsec distant to the star). Then, we can
calculate the degree of polarization ($DoLP$) within the stellar halo and its angle of linear polarization
($AoLP$):
\begin{eqnarray}
    {DoLP} &=& PI/I\,,\\
    {AoLP} &=& \frac{1}{2} {\arctan \left( U/Q\right)}\,.\label{equ:AoLP}
\end{eqnarray}

We found that the light carried in the stellar halo is polarized with a $DoLP$ of $0.46\pm0.08\%$ and with an
$AoLP$ of $98\pm8\,$deg East of North.  Because direct stellar light is expected to be entirely unpolarized, a
measurement of polarization from the isolated stellar halo indicates that there is an unresolved polarized
signal within the star-centered point spread function (PSF) of the observation.  This is typically attributed
to parts of the observed light scattering off interstellar dust within the line-of-sight, or due to an
unresolved disk component close to the star \citep[e.g.][]{ Keppler2018,vanHolstein2020}.  The measured $AoLP$
of the unresolved polarization is perpendicular to the PA of the outer disk, consistent with a co-planar inner
disk component.  This is further consistent with the continuous disk around MP~Mus, extending close to the
central star as observed with ALMA for gas and dust.  We subtracted the respective measured unresolved
polarization from the $Q$- and $U$-components before calculating $Q_\phi$.

\begin{figure*}
    \centering
    \includegraphics[width=\hsize]{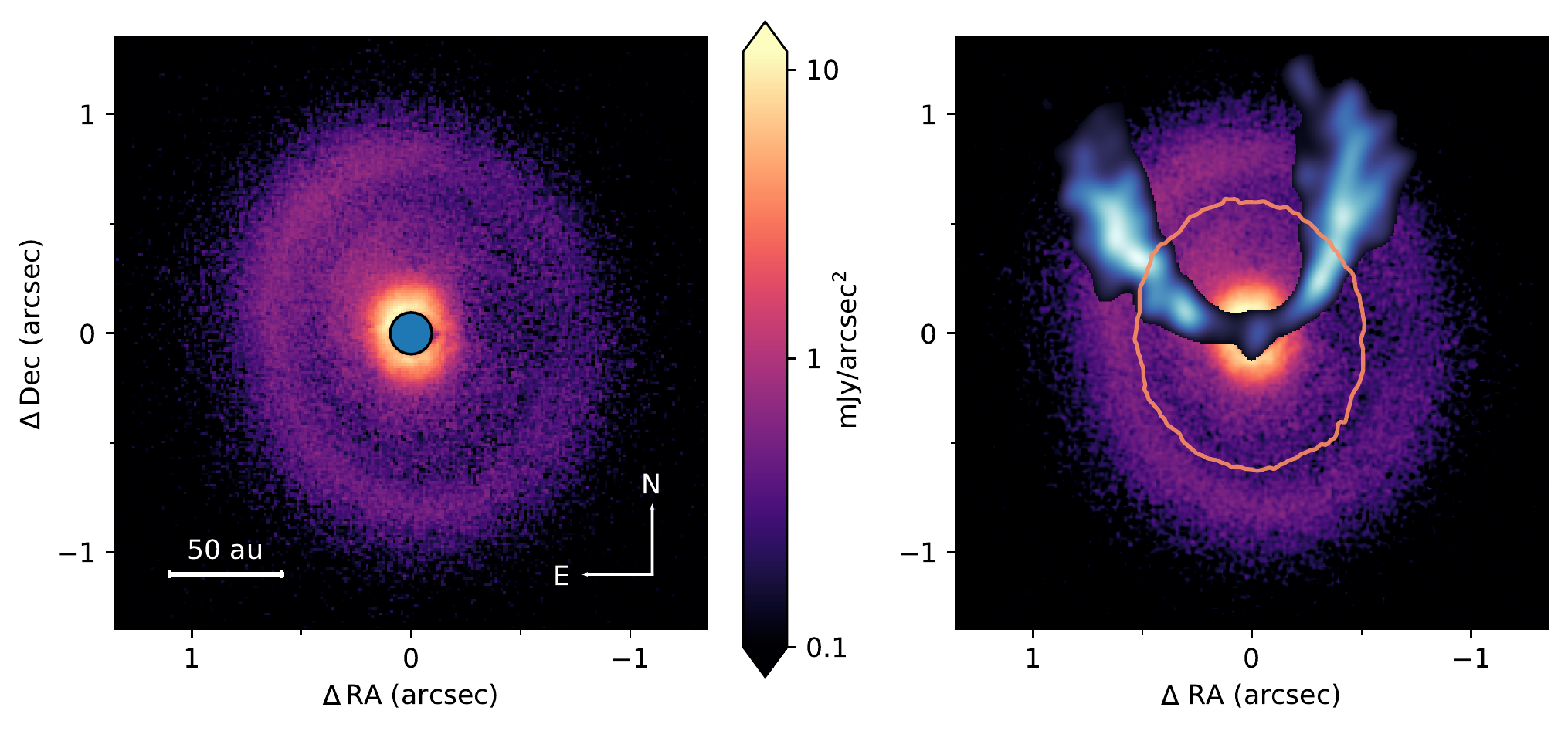}
    \includegraphics[width=\hsize]{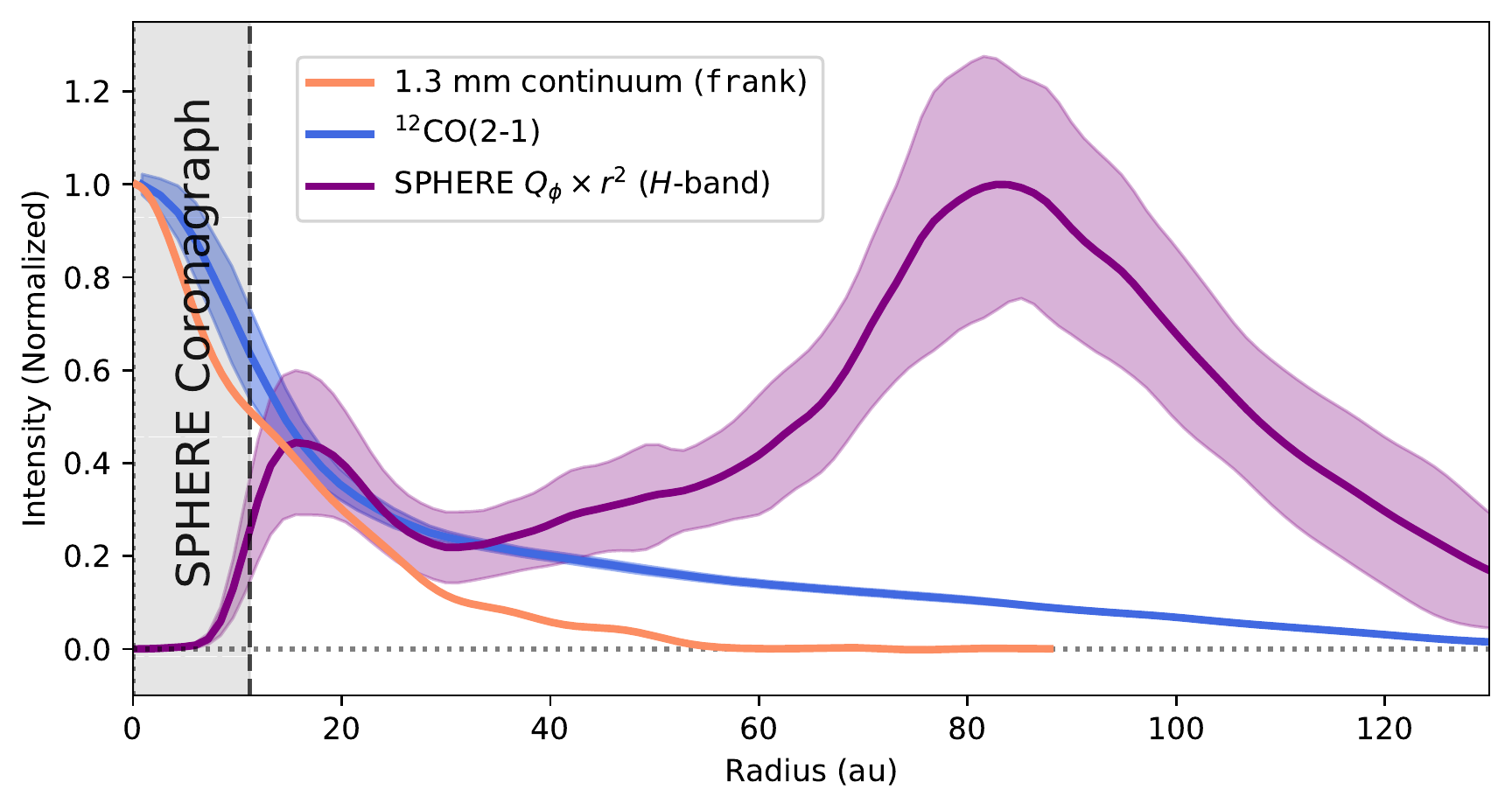}
    \caption{Comparison of ALMA and SPHERE observations of MP~Mus. \emph{Top left}: $Q_\phi$-image in $H$-band
      ($\lambda_{\rm obs}=1.6\,\mu$m) taken with SPHERE/IRDIS. The colormap is shown in logarithmic
      stretch. The central blue circle masks the area covered by the coronagraph of the observation. \emph{Top
        right}: Comparison of the 1.3\,mm continuum emission (orange contour), the SPHERE $Q_\phi$ $H$-band
      data (purple), and one of the $^{12}$CO (2-1) channels (blue). The 1.3\,mm contour corresponds to the
      5\,RMS level, while only data above the corresponding 3\,RMS levels are shown for the ${12}$CO (2-1) and
      scattered light observations. \emph{Bottom}: Comparison of the radial profiles of the SPHERE
      $Q_\phi \times r^2$ in $H$-band (purple line), the 1.3\,mm continuum (orange line), and the $^{12}$CO
      (2-1) emission (blue line). The SPHERE coronagraph is also shown.}
    \label{fig:SPHERE_results}
\end{figure*}

Figure~\ref{fig:SPHERE_results} shows the resulting $Q_\phi$-image of MP~Mus in $H$-band in logarithmic stretch
and the comparison with the continuum and $^{12}$CO (2-1) from ALMA. We used the flux frames of the IRDIS
observation to convert the observed polarized flux from counts to Jy/arcsec$^2$, assuming an $H$-band
magnitude of $7.64\pm0.02$ \citep[2MASS,][]{Cutri2003}. The final product is very similar to the presented
image in \citet{Avenhaus2018}. Figure~\ref{fig:SPHERE_results} also compares the radial profile of
$Q_\phi \times r^{2}$ in $H$-band with those of the ALMA 1.3\,mm and the $^{12}$CO (2-1) zero-th moment. The
SPHERE image show a bright inner region (the inner $\sim$20\,au are attenuated due to the SPHERE coronagraph)
and a decrease in intensity from 20 to 30\,au. The emission then increases slowly up to 60\,au and then rises
faster in an outer ring centered at $\sim$ 80\,au, dropping below the noise level at 130\,au, similar to the
extent of the $^{12}$CO (2-1). The slight asymmetry of this outer ring suggests that the near side of the disk
is to the east, the far side is west.  This is consistent with promoted forward-scattering
\citep[e.g.][]{Stolker2016}, as the near side of the disk appears brighter than the far side (by a factor of
$\sim1.5$ on average).

\citet{Avenhaus2018} mentioned an azimuthally-localized intensity decrease in the western half of the disk
(P.A.$\sim270^\circ$). This feature is also present in our image. Its location at the far side of the disk,
where signal-to-noise is lowest, makes it a very tentative detection. We further find that the disk's
brightest part is to the northeast of the star.

\section{Discussion}\label{sec:discussion}

\subsection{Comparison with stellar evolutionary models}\label{sec:comparison_evoltracks}

The stellar mass $M_{*,\rm dyn}=1.30\pm0.08\,M_\odot$ derived from the disk rotation
(Sec.~\ref{sec:dynamical_mass}) is independent of theoretical isochrones and evolutionary models, and offers
an interesting comparison with such models.

\begin{figure}
  \centering
  \includegraphics[width=\hsize]{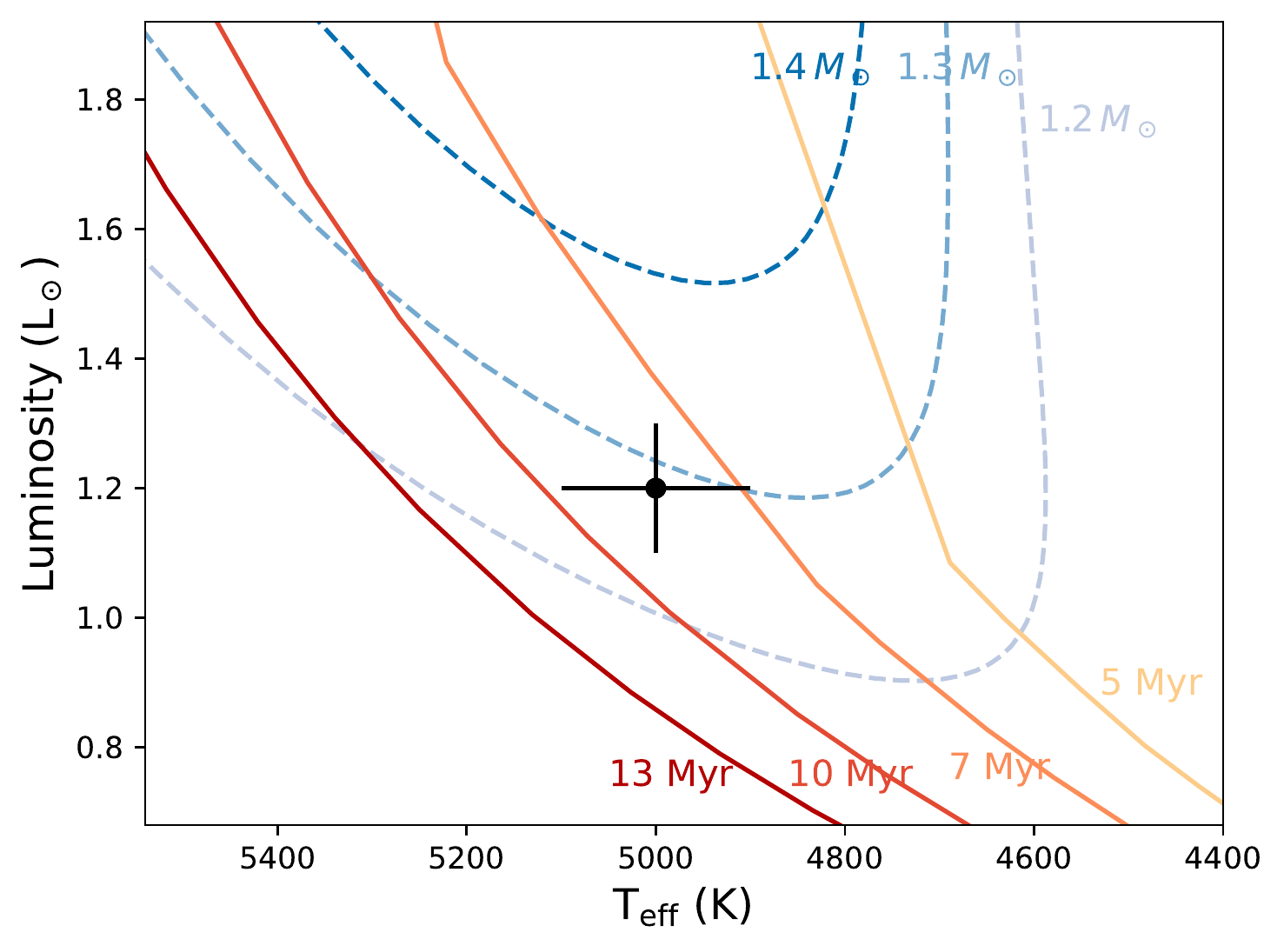}
  \caption{Location of MP~Mus (black dot) in the HR diagram. MIST mass tracks (1.2, 1.3, and 1.4\,$M_\odot$)
    and isochrones (5, 7, 10, and 13\,Myr) are also shown \citep{Dotter2016, Choi2016}.}\label{fig:MP_Mus_HR}
\end{figure}

MP~Mus is classified as a K1V star \citep{Mamajek2002}, corresponding to $T_{\rm eff}$ values between 4900 and
5100\,K in the spectral type (SpT)-$T_{\rm eff}$ relations of \citet{Kenyon1995} and \citet{Pecaut2013}. This
temperature is also in agreement with the 5110\,K value derived in the Gaia DR3 \citep{GaiaDR3}.  Its
interstellar extinction $A_V$ has been measured in the range of 0.2-0.7\,mag
\citep{Mamajek2002,Cortes2009}. More recently, \citet{AsensioTorres2021} derived values of 4600\,K and
0.8\,mag for this system from fitting the optical/near-IR SED, although this $T_{\rm eff}$ value appears a bit
too low for a K1 star based on the formerly mentioned SpT-$T_{\rm eff}$ tables. We perform a similar process
and fit the Tycho and 2MASS photometry using the BT-Settl photospheres \citep{Allard2011,Allard2012} with
$T_{\rm eff}$ values between 4500 to 5500 in 100\,K steps, $A_V$ values from 0 to 1\,mag in 0.1\,mag steps,
and stellar luminosity values between 0.8-1.6\,$L_\odot$ in steps of 0.1\,$L_\odot$. Adopting a distance of
d=98\,pc \citep{GaiaDR3} we derive values of $T_{\rm eff}$=4900\,K, $A_V$=0.6\,mag, and
$L_*$=1.3\,$L_\odot$. Such a $T_{\rm eff}$ value is higher than the effective temperature in
\citet{AsensioTorres2021} and more comparable to those adopted in earlier studies. Alternatively, we also
normalize 4900 and 5100\,K photospheres to the observed 2MASS photometry using $A_V$ values of 0.2 and
0.7\,mag, which yields stellar luminosities between 1.1-1.3\,$L_\odot$ (compatible with previous values when
corrected for the updated distance from Gaia). We thus adopt $T_{\rm eff} =5000 \pm 100$\,K and
$L_* =1.2\pm0.1$\,$L_\odot$ for MP~Mus.  Figure~\ref{fig:MP_Mus_HR} shows its location in the corresponding HR
diagram using the MESA Isochrones and Stellar Tracks \citep[MIST,][]{Dotter2016, Choi2016}. This results in a
stellar mass and age of $\sim$1.3\,$M_\odot$ and 7-10\,Myr for the system. MP~Mus is a confirmed member of the
old $\epsilon$~Cha association based on its kinematic properties
\citep[e.g.,][]{Murphy2013,Dickson-Vandervelde2021}. \citet{Murphy2013} assigned a 3-5\,Myr age to that region
(younger than our estimate), but the re-analysis of $\epsilon$ Cha by \citet{Dickson-Vandervelde2021} including
Gaia data yielded an age of $5^{+3}_{-2}$\,Myr that is compatible with our results. A lower $T_{\rm eff}$ such
as the one proposed in \citet{AsensioTorres2021} would result in a younger (but still compatible) age for
MP~Mus, although the dynamical stellar mass of $\sim$1.3\,$M_\odot$ derived from the ALMA
observations appears more difficult to reconcile with this lower $T_{\rm eff}$ value.

The mass derived from the MIST tracks also appears fully compatible with the dynamical mass estimate. Given
the good agreement between various recent pre-main sequence evolutionary models \citep[][]{SimonToraskar2017},
this result is not restricted to these tracks.  Studies comparing dynamical stellar masses with
  predictions from evolutionary models without magnetic fields typically find the latter to underestimate masses by a
  significant amount, especially for low-mass stars \citep[by 30\,\%-80\,\% for stellar
  masses $<$1.4\,$M_\odot$, e.g.,][]{Simon2019, Pegues2021}. In contrast, evolutionary tracks including
  stellar magnetic fields \citep[e.g.,][]{Feiden2016} yield more compatible results with dynamical mass
  estimates. This discrepancy is usually attributed to starspots \citep[e.g.,][]{Pegues2021, Flores2022},
  which become more relevant for late-type stars. Given the relatively high mass of MP~Mus compared to the
  samples in the aforementioned works, the agreement between the dynamical mass estimate and the evolutionary
  tracks without magnetic fields used in our study is likely an indication that the effect of starspots in this system is
  moderate/negligible. We also notice that these works focused on younger sources (mostly in
  Taurus and Ophiuchus, with estimated ages of 1-2\,Myr). Similar comparisons for larger
  samples of stars covering a range of masses and ages could help to further inform theoretical models of
  early stellar evolution.

\subsection{Grain growth and optically thick emission in MP~Mus}\label{sec:grain_growth}

Assuming optically thin emission and a sufficiently warm dust temperature for the emission to be in the
Rayleigh-Jeans regime, $\alpha_{\rm mm}$ relates to the power law index of the dust opacity,
$\kappa_\nu \propto \nu^\beta$, via $\alpha_{\rm mm}=2+\beta$.  $\beta$ is sensitive to the maximum grain size
($a_{\rm max}$) of the dust size distribution \citep[e.g.][]{Natta2004b, Draine2006, Testi2014}: dust
distributions with maximum grain sizes around or larger than 1\,mm have opacity power-law indices
$\beta\lesssim 1$, while smaller grains have higher $\beta$ values (ISM-like grains show $\beta \sim 1.7$). The
spectral indices measured in Sec.~\ref{sec:spectral_indices} then translate to average $\beta$ values of
$\sim$0.1-0.4 and suggest the presence of mm-sized grains in MP~Mus. Our results are very similar to those of
\citet{Cortes2009}, who found the mm spectral index of MP~Mus between 3\,mm and 12\,cm to be
$\alpha=2.4\pm0.1$, and first indicated the presence of large grains in the disk. Similar $\alpha_{\rm mm}$
values are found in disks in younger star-forming regions such as Taurus, Ophiuchus, Lupus, and Chamaeleon~I
\citep[][]{Ricci2010_Taurus,Ricci2010_Ophiuchus,Ribas2017,Tazzari2021_lupus}, but also in other evolved disks
such as TW~Hya, HD~98800B or TWA~3\,\citep{Macias2021, Ribas2018, Czekala2021}. This implies that, despite
grain drift and evolution, protoplanetary disks can retain a population of large grains during most of their
lifetimes.

The spectral index map in Sec.~\ref{fig:alpha_map_and_profile} provides additional spatial information about
the disk, and two of its features are worth discussing. First, $\alpha_{\rm 1.3 - 2.2\,mm}$ increases with
radius, similar to many other sources observed at multiple (sub)mm wavelengths with enough angular resolution
\citep[e.g.,][]{Perez2015,Carrasco2016,Tazzari2016,Dent2019,Macias2019,Macias2021}.  This is attributed to a
combination of higher optical depths in the inner regions and to the inward drift of large grains due to gas
drag \citep[][]{Weidenschilling1977}. Thus, this can be interpreted as additional evidence of grain growth and
radial drift in the disk.  It is also worth noticing that the disk radii at 1.3\,mm and 2.2\,mm are
  similar within the uncertainties. Although this interpretation is limited by the moderate angular resolution
  of the 2.2\,mm observations ($\sim$20\,au), if both radii are indeed similar this could suggest the presence
  of some mechanism preventing strong dust radial drift in the disk, since it would otherwise appear more
  compact at 2.2\,mm than at 1.3\,mm.

Maybe more interesting is the fact that MP~Mus displays values $\alpha_{\rm 1.3 - 2.2\,mm}< 2$ in its inner
region ($r<$30\,au). Such values are below the spectral index of black body radiation, and they have been
traditionally interpreted as indicators of additional emission mechanisms such as free-free radiation from
ionized photoevaporative winds or stellar chromospheric activity
\citep[e.g.,][]{MacGregor2015,Macias2016}. However, these processes usually become significant at wavelengths
longer than those considered here. \citet{Cortes2009} also observed MP~Mus at cm wavelengths and obtained only
upper limits at 3\,cm and 6\,cm, suggesting a negligible contribution from a possible stellar
wind. Alternatively, $\alpha_{\rm mm}<2$ values can also arise from optically thick emission from dust grains
with high albedo \citep{Miyake1993,Liu2019,Zhu2019}.  Recently, ALMA has shown that spectral indices below 2
in the inner regions of disks are not unusual \citep[e.g.][]{Huang2018,Dent2019}, and an integrated
$\alpha_{\rm mm}$ value below 2 was also found in the circumbinary disk of the TWA~3 triple system
\citep{Czekala2021}. The spectral index map of MP~Mus is one more example of these cases, adding observational
support to the idea that at least part of the emission from protoplanetary disks is optically thick even at
$\sim$2\,mm, and that dust scattering needs to be considered at these wavelengths. In turn, this is yet
another indication that dust masses derived from mm surveys may be underestimated, and explains why SED
modeling of disks accounting for the disk structure and dust scattering results in systematically higher dust
masses \citep{Ballering2019,Ribas2020,Rilinger2023}. The fact that these low $\alpha_{\rm mm}$
values are also found on evolved disks such as TW~Hya or MP~Mus shows that this effect could be significant
throughout most of the disk lifetime.

\subsection{Lack of mm substructures and comparison with TW~Hya}\label{sec:no_substructures}

The high angular resolution view of MP~Mus presents a stark contrast with respect to most protoplanetary disks
imaged at similar spatial resolutions to date: while the majority of disks display some substructures in high
angular resolution mm observations (i.e. gaps, rings, spiral arms, asymmetries), MP~Mus shows a smooth disk
down to a 4\,au resolution, with the possible exception of a barely resolved outer ring at $\sim$50\,au. There
are also no signs of an inner cavity, in agreement with the $AoLP$ derived from the SPHERE observations
(Sec.~\ref{sec:sphere_results}) as well as with previous modeling of its SED which suggested that the disk
extends down to the dust sublimation radius \citep{Cortes2009}. The featureless appearance is even more
puzzling when we consider its age (7-10\,Myr), which is quite older than typical disk lifetimes. Without
substructures acting as dust traps \citep[e.g.][]{Pinilla2012, Zhu2014}, the gas drag would quickly deplete
the disk of large grains on timescales much shorter than disk lifetimes \citep[][]{Weidenschilling1977}, so
one of the following scenarios must be true for MP~Mus: either its dust population comprises small
($<100$\,$\mu$m) dust grains only that are weakly affected by radial drift, or there are undetected/unresolved
substructures in the disk that are stopping the inward migration of large grains.

In the first scenario, a population of smaller grains would have both a lower opacity at mm wavelengths and a
steeper opacity power law index \citep[e.g.][]{Natta2004b,Testi2014,Tazzari2016}. A significantly higher dust
mass would be needed to both account for the observed fluxes and to maintain a spectral index of 2-2.5 in most
of the disk, which in this case would be mostly due to optically thick emission. Although this explanation
cannot be ruled out with the current data, it seems unlikely that the actual dust mass would be much higher
than the measured value, as it would imply that MP~Mus was originally very massive (possibly above the disk
instability limit). Moreover, the disk size at mm wavelengths and those of the gas and scattered light
observations would be difficult to reconcile with a population of small grains only.  Additional
high-resolution observations at other mm wavelengths covering a broader wavelength range are needed to perform
a detailed modeling of the dust population in the disk \citep[e.g.,][]{Macias2021} and to further investigate
this possibility.

\begin{figure}
  \centering
  \includegraphics[width=\hsize]{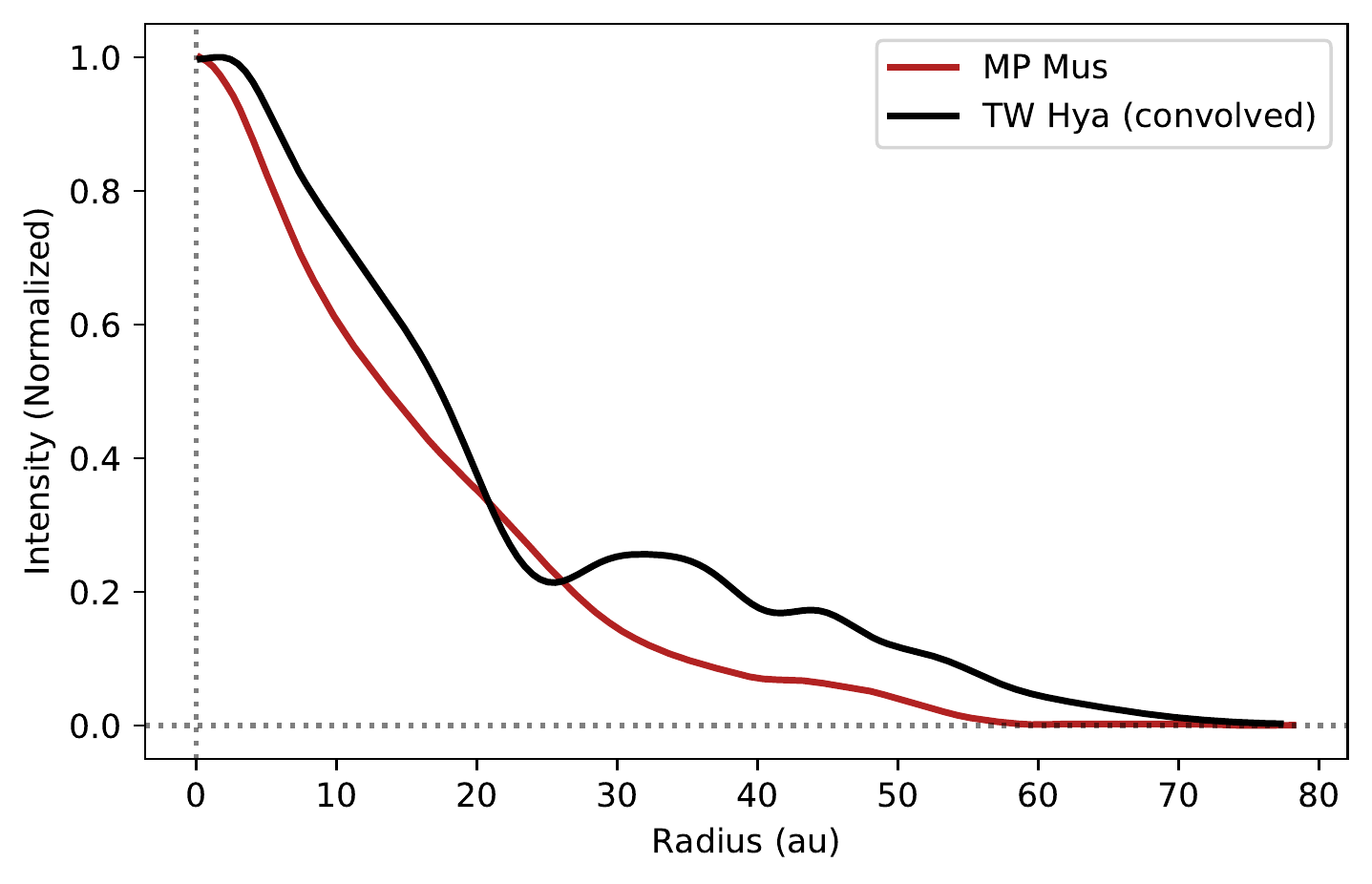}
  \caption{Comparison of the radial profiles of MP~Mus (red) and TW~Hya (black) at 1.3\,mm. The TW~Hya data
    were taken from \citet{Macias2021}, and convolved with a 0.08\arcsec Gaussian to match the resolution and
    distance of the MP~Mus observations. The inner cavity and two of the known gaps gaps of TW~Hya (at 25 and
    45 au) are still visible in the case of TW~Hya, in contrast with the smooth profile of
    MP~Mus.}\label{fig:TWHya_comparison}
\end{figure}

The second explanation for the survival of large grains in an evolved disk with apparently no substructures is
that this conclusion is limited by the optical depth and/or angular resolution of the observations, i.e.,
there are structures in the disk but they remain undetected or unresolved with the current
observations. Structures narrower than 5\,au have been found in some nearby disks such as TW~Hya, HD~169142
and V4046~Sgr \citep{Andrews2016, Perez2019, MartinezBrunner2022}, although the last two also displayed much
broader gaps in the dust distribution (and V4046~Sgr is a circumbinary disk). TW~Hya probably provides the
best comparison given its similarities with MP~Mus: high angular resolution observations revealed a system of
concentric rings and gaps with 1-5\,au widths and varying amplitudes \citep{Andrews2016}. These authors argued
that such structures may be common in disks, playing a fundamental role in their evolution and in the planet
formation process. Figure~\ref{fig:TWHya_comparison} compares the radial profiles of MP~Mus and TW~Hya at
1.3\,mm from \citet{Macias2021} after degrading the angular resolution of the latter to match the $\sim$5\,au
of the MP~Mus data. This comparison demonstrates that both profiles are significantly different, and it is
still possible to identify several features in TW~Hya at this resolution such as a flat profile in the inner
region (caused by the central cavity) and two gaps at 25 and 40\,au. On the other hand, a number of the known
substructures in TW~Hya are no longer visible in the profile, as could be the case for MP~Mus. We note that
gas rings with radial widths much smaller than their vertical extent are unstable \citep[][]{Ono2016,
  Dullemond2018}, which limits how packed substructures can be. Given the flat appearance of MP~Mus, realistic
values for $h/r$ are likely in the range of 0.03-0.07, which could allow for the presence of rings in the
outer regions that are narrow/close enough to remain unresolved.  This consideration is also relevant for the
comparison with TW~Hya: the stellar mass of MP~Mus is $\sim$twice that of TW~Hya \citep[1.3\,$M_\odot$ for
MP~Mus, and $\sim$0.6\,$M_\odot$ in the case of TW~Hya][]{Sokal2018}, so its scale height is expected to be
smaller (by a factor of $\sim\sqrt{2}$ when ignoring the higher stellar luminosity of MP~Mus). As a result,
MP~Mus may host narrower rings than TW~Hya. In addition, \citet{Xu2022a, Xu2022b} showed that within one
  broad gas pressure bump, two radial dust rings separated by a distance comparable or less than $h$ may form,
  assisted by the dust back reaction to gas. This further promotes the possibility of having packed dust rings
  currently unresolved.  Finally, high optical depths also limit our ability to identify gaps/rings in the
system. As suggested by the map of spectral index, the inner 30\,au of the disk appear to be optically thick,
displaying $\alpha<2$ (Fig.~\ref{fig:alpha_map_and_profile}). It is thus possible that the observed 1.3\,mm
continuum emission in these regions is coming from an elevated surface, preventing the detection of
substructures in the midplane.

Overall, despite the high angular resolution and structureless appearance of MP~Mus, the current data cannot
discard small ($<$4\,au) rings especially in the inner regions, or even larger ones if the disk is sufficiently optically
thick. If undetected substructures are the explanation for the long-lived disk around MP~Mus, then this system
lends further support to the idea that small rings may be a common feature in disks. As proposed by
\citet{Tripathi2017, Tripathi2018, Andrews2018}, if some of these rings are (partially) optically thick, they
could help to explain the observed relation between the radii and mm luminosity of disks. Evolutionary disk
models by \citet{Toci2021} also showed that the gas-to-dust radii ratios ($R_{\rm CO}$/$R_{\rm dust}$) become
$>$5 in very short timescales ($<$1\,Myr) in the absence of mechanisms preventing radial drift, in clear
conflict with the $R_{\rm CO}$/$R_{\rm dust}\sim 2$ value found in MP~Mus at 7-10\,Myr. Given its proximity,
ALMA observations of MP~Mus at even better angular resolutions and longer wavelengths would aid in explaining
the long lifetime of the system and exploring the frequency and properties of small structures in disks.

\subsection{Upper limits to the presence of planets}\label{sec:upper_planets}

\begin{figure}
  \centering
  \includegraphics[width=\hsize]{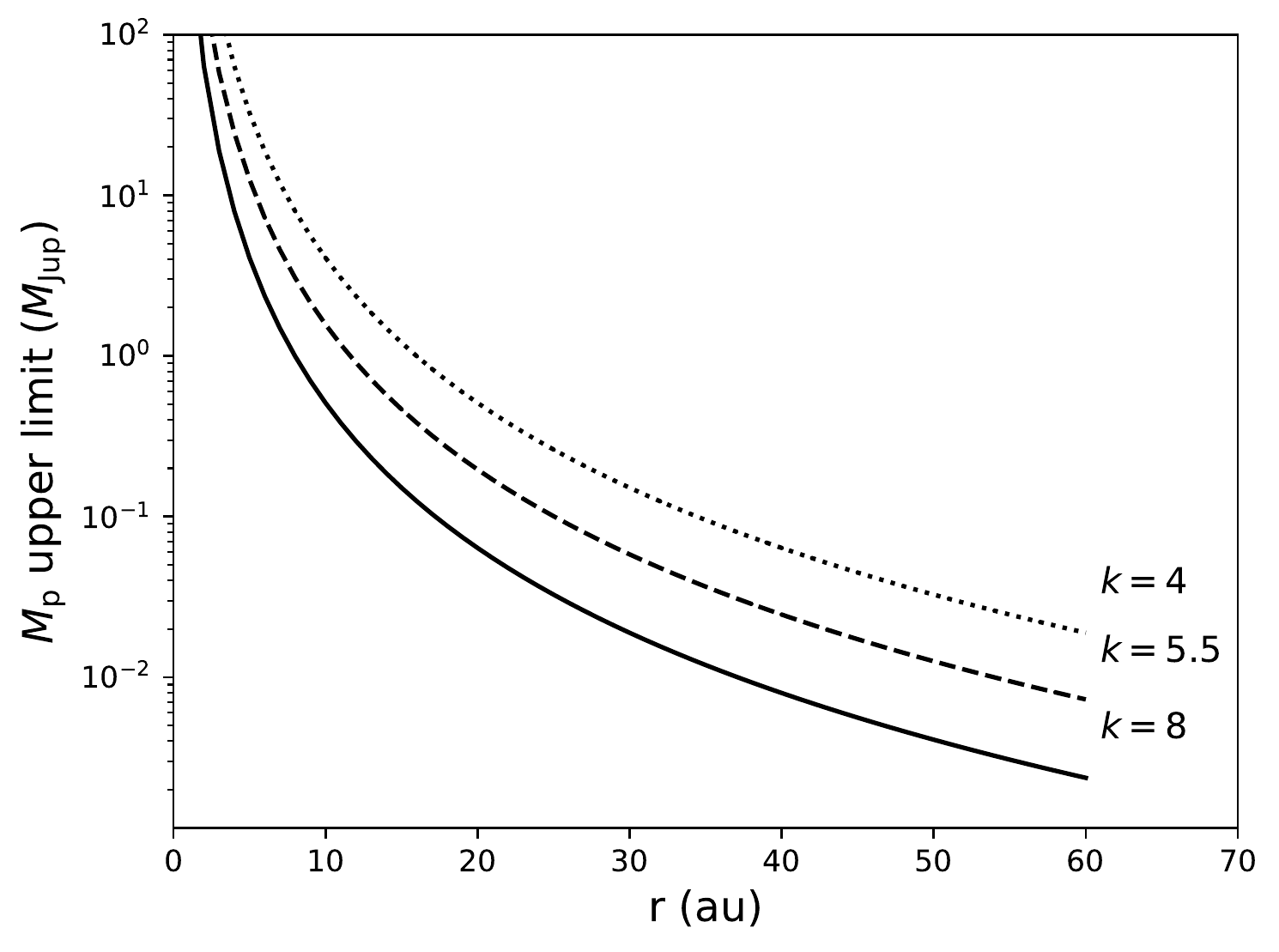}
  \caption{Upper limits to the presence of planets in the disk around MP~Mus as a function of radius. These
    upper limits are derived using equation~\ref{eq:Mp} \citep{Lodato2019} and the fact that no gaps with
    width $\triangle \geq 5$\,au are detected in the disk. Three different values are used for the proportionality
    constant ($k$) linking the planet Hill radius and the corresponding gap width.}\label{fig:Mp_upperlims}
\end{figure}

Planets with enough mass are expected to open gaps in the gas and dust distribution
\citep[e.g.,][]{Crida2006, Zhu2011, Dong2017, Dong2018}, although the properties of these gaps are highly
dependent on the conditions in the disk.  Deriving planetary masses from gap widths and contrasts requires
detailed hydrodynamic models including many (uncertain) free parameters and is a quite degenerate process, but
comparisons with results from different models provide some constraints on the maximum mass of planets in the
MP~Mus system based on the absence of visible substructures larger than $>$4\,au in dust distribution. Here,
we use the scaling relation between the planet Hill sphere ($R_{\rm H}$) and width of the gap ($\triangle$,
defined as the separation between the minimum brightness in a gap and the maximum brightness of the
corresponding external ring) following \citet{Lodato2019}. For disks with viscosity values
$\alpha\lesssim 0.01$, these two quantities are related following:
\begin{equation}\label{eq:Mp}
\triangle = k R_{\rm H} = k \biggl(\frac{M_{\rm p}}{3 M_*}\biggr)^{1/3}r,
\end{equation}
where $M_{\rm p}$ is the mass of the planet, $M_{\rm *}$ is the mass of the star, $r$ is the orbital radius of
the planet, and $k$ is a proportionality constant that depends on disk properties and ranges between $\sim$4-8
for gaps observed in the mm \citep[see][and references therein, where they adopt a value of
$k$=5.5]{Lodato2019}. Assuming that we would have resolved widths $\triangle \geq 4$\,au and the derived
1.30\,$M_\odot$ mass for MP~Mus, Fig.~\ref{fig:Mp_upperlims} shows the corresponding upper limits for
$M_{\rm p}$ at different radii for $k$=4, 5.5, and 8. Based on this simple relation and further assuming that
each gap is carved by a single planet, the lack of resolved gaps in these observations would imply that there
are no planets more massive than 0.5-4\,$M_{\rm Jup}$ at orbital radii $r > 10$\,au (depending on the adopted
value of $k$), and these constraints decrease to $0.05-0.5$\,$M_{\rm Jup}$ and 2\,$M_\oplus - $0.06\,
$M_{\rm Jup}$ for $r > 20$\,au and $r > 40$\,au, respectively. We emphasize that this analysis assumes that
any sufficiently large gap carved by planets would be visible in the 1.3\,mm observations, but the high
optical depth within the inner 30\,au suggested by their $\alpha<2$ spectral index could hide such gaps (see
Sec.~\ref{sec:no_substructures}). Therefore, the derived limits are likely underestimated, especially in the
inner regions of the disk. Also, these limits only apply for $r \leq 60$\,au since that is the radial extent
of the mm dust continuum emission.

The 1.3\,mm radial profile obtained with \emph{frank} shows a shallow plateau between 30-40\,au, which could
be interpreted as a low-contrast gap. Although it is not clear that this structure is an actual gap in the
dust density distribution of the disk, here we also use Eq.~\ref{eq:Mp} to calculate the mass of a
hypothetical planet that could carve such a gap at that location. Assuming that the planet is at the center of
the plateau ($\sim$35\,au) and given the location of the outer bump ($\sim$45\,au), the gap width would be
$\triangle$=10\,au, and the corresponding planet mass is 0.2\,$M_{\rm Jup}$, 0.6\,$M_{\rm Jup}$, and
1.5\,$M_{\rm Jup}$ for a value of $k$ of 8, 5.5, and 4, respectively.
  
On the other hand, the drop in scattered light signal between 30-80\,au seen in the radial profile of the
scattered light (see Sec.~\ref{sec:panchromatic_view}) may indicate the presence of one or more massive
planets outside the continuum emission, and studies based on scattered light data have placed different upper
limits to their masses. \citet{Wolff2016} observed MP~Mus with GPI and could reject the presence of a
3\,$M_{\rm Jup}$ planet outside of 40\,au with a 90\,\% confidence for a 7\,Myr disk. Similarly, SPHERE
observations of MP~Mus allowed \citet{AsensioTorres2021} to place upper limits on the masses of possible
companions\footnote{The planetary mass upper limits in \citet{AsensioTorres2021} were derived assuming a
  3\,Myr age for MP~Mus. If the system is older as suggested by our analysis and other studies, then the upper
  limits would be higher.}, now reaching a 5-$\sigma$ limit of $\sim$2.5-3\,\,$M_{\rm Jup}$ at $r
\geq$50\,au. \citet{AsensioTorres2021} also proposed that a planet located at 55\,au may be the origin of this
decrease in scattered light intensity. However, the lack of a wide gap at this location in the ALMA continuum
observations suggests that the lower intensity of scattered light between 30-80\,au is not due to an actual
gap in the disk surface density (see Sec.~\ref{sec:panchromatic_view}).  Recently, the GAPlanetS survey
\citep{Follete2022} targeted several sources in H$_\alpha$ searching for accreting protoplanets in fourteen
disks, but did not identify any potential candidate in the disk around MP~Mus. If the lack of any clear
rings/gaps in the 1.3\,mm continuum is real and not due to the high optical depth of the disk in the inner
regions, then this work and previous studies analyzing scattered light observations appear to rule out the
presence of planets more massive than $3\,M_{\rm Jup}$ at radii $r > 10-15$\,au. If that is the case, the
exoplanet population of a hypothetical planetary system around MP~Mus may not be very different from that of
the Solar System, making it an ideal laboratory for further planet formation studies given its proximity and
age.

Finally, the mm observations in this work also place some constraints on the presence of circumplanetary disks
outside of the dust disk ($r>$60\,au). The RMS of the 1.3\,mm continuum image is 19\,$\mu$Jy/beam, and thus we
can reject the presence of point-like sources with fluxes $\geq 90$\,$\mu$Jy in the dust-free region at a
5\,$\sigma$ confidence. There are two $\sim 4$\,$\sigma$ peaks at angular separations 1.4'' NE and 1.6'' SE
from the source center ($\sim$140 and 160\,au projected orbital radii) which lie outside the scattered light
outer ring, but these are likely associated with noise. Converting these limits to disk masses involves many
unknown quantities \citep[see][]{Isella2014, Isella2019}, but we can compare the sensitivity limit of these
data to the flux of the circumplanetary disk recently detected around PDS~70c
\citep[][]{Isella2019,Benisty2021}: considering its 86\,$\mu$Jy flux at 855\,$\mu$m, an spectral index of 2.3,
and the closer distance of MP~Mus, a similar circumplanetary disk would have a flux of 42\,$\mu$Jy at 1.3\,mm
and would therefore remain undetected with our observations.

\subsection{The panchromatic view of MP~Mus}\label{sec:panchromatic_view}

The ALMA observations of MP~Mus reveal a smooth disk in the mm continuum with a radius of 60\,au, and a
gaseous disk that extends further out to 130\,au. In contrast, the scattered light images probing small dust
grains show a different morphology, with a bright inner region and a significant drop in flux between
30-80\,au \citep[][]{Cortes2009,Schneider2014,Wolff2016,Avenhaus2018}. In particular, SPHERE observations of
MP~Mus show that: (1) the strongest signal in scattered light arises from the inner 25\,au, (2) there is a
discontinuity in the brightness at 60\,au, (3) there is an outer ring at 80\,au, and (4) the disk extends up
to 125\,au. Many of these coincide with features seen in the ALMA images (see Fig.~\ref{fig:SPHERE_results}
for a comparison of the mm continuum, $^{12}$CO, and scattered light observations and radial
profiles). Regarding the inner regions, the 1.3\,mm continuum emission also arises mostly from the inner
25-30\,au, the slopes of the radial profiles of both continuum and $^{12}$CO emission changes at this radius
(Fig.~\ref{fig:continuum_and_12CO_radial_profiles}), and the emission has a spectral index
$\alpha_{\rm mm} <2$ within this radius (Fig.~\ref{fig:alpha_map_and_profile}). The discontinuity observed at
60\,au in scattered light coincides with the outer radius of the mm continuum emission, and the extent of the
gaseous component of the disk is in great agreement with the one inferred from the scattered light
data. Overall, this matches predictions from the radial drift of dust grains: while small grains remain well
coupled to the gas and are thus co-located, larger grains migrate toward pressure maxima and accumulate in the
inner regions of the disk or localized pressure bumps \citep[e.g.,][]{Weidenschilling1977,Pinilla2012}.

The gap seen in scattered light between 30-80\,au is quite prominent \citep[it is the largest of all the
sources in][]{Avenhaus2018}, and two scenarios have been proposed to explain it, namely that it is a
real decrease in the disk density carved by one or multiple planets, and that it is shadow cast by the disk
itself \citep[][]{Wolff2016}. We can now revisit these explanations in the light of the new ALMA observations.

Planets are one of the leading explanations for the plethora of rings and gaps found in protoplanetary disks,
and the properties of many of these structures appear to support this explanation
\citep[e.g.,][]{Huang2018,Zhang2018,Perez2019}. In these cases, the gravitational influence from the planet
decreases the local gas density and induces a pressure bump outside of its orbit, which acts as a dust trap
for large (mm/cm-sized) grains. For MP~Mus, however, the lack of any mm emission in the outer ring visible in
scattered light implies that the ring is mostly devoid of large dust grains, suggesting that the gap in
scattered light is not a gap in the surface density opened by one or several planets. Although the
available observations do not reject planet masses below sub-Jovian values, the lack of companions discussed
in Sec.~\ref{sec:no_substructures} also supports the interpretation that the drop in scattered light at
30-80\,au is due to a different process.

A shadow cast by the disk inner rim \citep[e.g.][]{Dullemond2004} is also a plausible explanation for the
apparent gap in the scattered light observations, especially considering that MP~Mus shows a rather flat
structure. The possibility of a shadowed disk in the system was proposed by \citet{Cortes2009} based on its
low near/mid-IR excess (4-20\,$\mu$m) with respect to the median disk SED of Taurus, even before high-angular
resolution observations in scattered light were available. \citet{Dong2015} found that a puffed-up inner rim
with a sharp edge in the vertical direction could produce a pattern similar to that seen in MP~Mus (i.e., a
drop in scattered light emission at intermediate radii), although rims with more physically-motivated
structures did not produce such results. It is interesting to note that a steep increase in the SPHERE radial
profile occurs at 60\,au, matching the outer radius of the mm emission. Based on this and the
overall SPHERE radial profile, we propose that MP~Mus probably has a puffed-up inner disk that shadows radii
beyond 30\,au. However, the lack of large grains in the disk midplane at radii r$>$60\,au could result in a
less efficient cooling of the disk (and hence a warmer midplane), increasing in the disk scale height and
allowing the disk surface to grow out of the shadow at longer radii. Direct observational evidence for such an
effect (i.e., higher temperatures in the outer regions devoid of large grains) exist for the edge-on disk
Oph~163131 based on tomographic reconstruction of its temperature structure
\citep[][]{Flores2021,Villenave2022}.

We explore this idea by calculating a simple disk model for MP~Mus using the MCFOST code
\citep{MCFOST,MCFOST2}. We adopt the stellar and disk parameters derived in this study and include two
different dust populations: one for small grains (0.01\,$\mu$m - 10\,$\mu$m) extending from 0.1 to 130\,au,
and a more compact disk of larger grains (10\,$\mu$m - 1\,mm) from 0.1 to 60\,au. At 60\,au, the inter-phase
between the large and small grain disks results in an increase of the midplane temperature from 10 to
15\,K. For a vertically isothermal disk with a scale height $H=c_s/\Omega$, this change in temperature would
imply a local increase of $\sim$20\,\% in the scale height, which could expose the upper disk layers to
stellar radiation again. We emphasize that none of these numbers are to be considered as accurate estimates
for MP~Mus given the oversimplified model used, but they show that the change in the disk opacity at the end
of the mm continuum radius could lead to a local increase of the disk temperature at that location.  We do not
find signs of a colder disk at $r = 60-80$\,au in the $^{12}$CO (2-1) radial profile, which does not show any
significant feature at these radii down to 3\,K (the brightness temperature uncertainty at 70\,au). However,
the $^{12}$CO (2-1) emission arises from above the midplane and may not reflect the midplane temperature.
Unfortunately, the same analysis cannot be performed for optically thinner isotopologues such as $^{13}$CO
(2-1) or C$^{18}$O (2-1), since their emission only extends to $\lesssim$60\,au. Detailed physico-chemical
modeling of the gas and dust components of MP~Mus is needed to calculate a consistent temperature structure
for the disk and to determine the origin of the gap seen in the scattered light observations.

\section{Summary}\label{sec:summary}

We present new ALMA observations of the nearby, evolved protoplanetary disk around MP~Mus, including 0.89\,mm,
1.3\,mm, and 2.2\,mm continuum emission, as well as multiple gas emission lines. These data are the first
spatially resolved observations of the disk at these wavelengths and provide a wealth of new information of
this system. Our key results and findings are:

 \begin{itemize}

 \item The continuum emission shows a disk with no detected inner cavity and a radius of
   60$\pm$5\,au. Despite the high angular resolution of these data, the dust disk
   appears smooth down to 4\,au scales, making MP~Mus an interesting exception when compared with the plethora
   of substructures found in most disks observed at comparable resolutions, and an great counterpart to TW~Hya
   in particular.

 \item Based on the mm fluxes and using standard assumptions for the dust opacity and disk temperature, we
   derive a dust disk mass of $M_{\rm dust}=0.14_{-0.06}^{+0.11}$\,$M_{\rm Jup}$.

 \item The continuum spectral index between 1.3 and 2.2\,mm has a value $< 2$ for radii $r<30$\,au,
   indicative of optically thick emission from dust grains with a high albedo in this region.

 \item These observations yield detections of  $^{12}$CO (3-2), CS (7-6), HC$^{15}$N (4-3), and $^{13}$CO
   (3-2) in Band 7, $^{12}$CO (2-1), $^{13}$CO (2-1), and C$^{18}$O (2-1) in Band 6, as well as HC$_3$N
   (16-15), DCN (2-1), and DCO$^+$ (2-1) in Band 7.

 \item The $^{12}$CO (2-1) observations reveal a gaseous disk extending up to
     130$\pm$15\,au, a factor of $\sim$2 larger than the dust disk. Similar to the
   dust, no clear gaps are found.

 \item By fitting a Keplerian profile to the first moment of the $^{12}$CO (2-1), we derive a
   dynamical mass for MP~Mus of $1.30 \pm 0.08 M_\odot$. This value is consistent with predictions from
   theoretical stellar evolutionary models, which date the system at an age of 7-10\,Myr.

 \item By comparing the $^{13}$CO (2-1) and C$^{18}$O (2-1) fluxes with grids of disk models, we estimate the gas
   mass in the system to be $10^{-4}-10^{-3}$\,$M_\odot$, resulting in a global gas to dust ratio of 1-10.

 \item A comparison of these data with previous scattered light observations shows that small grains and
   the gas are co-located while larger grains concentrate in the inner regions, in line with expectations from
   dust radial drift.

 \item From the scattered light observations, we derive an angle of linear polarization that speaks for
     disk material inside the stellar PSF, co-planar with the outer disk. This is in agreement with the
     expectations from a disk extending inward to regions close to the star (within ~0.05\arcsec).
       
 \item The survival of large grains in a gas-rich disk is surprising for such an evolved system, especially
   considering the lack of substructures in the continuum emission. This suggests that structures preventing
   the radial drift of large grains may be present in the disk but not visible due to a high optical of the
   emission at 1.3\,mm, or they may be smaller than the resolution limit of these observations (4\,au).

 \item Based on scaling relations between the planet mass and the gap widths and assuming that substructures
   are not hidden by high optical depths, we use the lack of any clearly resolved gap to place upper limits to
   the mass of possible planets in the disk. We find that no planets more massive than
   $\sim$0.5-4\,$M_{\rm Jup}$, $0.05-0.5$\,$M_{\rm Jup}$, and 2\,$M_\oplus-$0.06\, $M_{\rm Jup}$ exist at
   radii $r>$10, 20, and 40\,au, respectively. Using a similar approach, if the tentative plateau seen at 30 -
   40\,au is caused by a planet, it would have a mass $\sim 0.2-1.5\,M_{\rm Jup}$.

 \item The scattered light observations also revealed a drop in intensity between 30 and 80\,au, and an
     outer ring from 80 to 130\,au. The lack of mm emission from this outer ring suggests this drop in
     scattered light emission is probably not an actual gap in the disk surface density due to planets, since
     such a gap would trap large grains in the ring. Instead, the data appear more consistent with this
     feature being a shadow, cast between 30-80\,au by a puffed-up inner rim. The rapid increase of scattered
     light signal at radii $>$ 60\,au may be explained by the lack of large dust grains at these locations,
     which could result in a warmer disk. This would increase the disk scale height and expose the disk
     surface to stellar radiation at longer radii, explaining the outer ring visible in scattered light.
   
  \end{itemize}
  
 Given its nearby location, age, and properties, MP~Mus is an optimal target to study many aspects of
 protoplanetary disks, a great laboratory to probe the chemistry of planet formation, an interesting
 counterpart to the TW~Hya system and, possibly, the closest analog to the young Solar System. Because of all
 these factors, MP~Mus may be one of the most promising individual sources to advance our understanding of
 planet formation.

\begin{acknowledgements}
  We thank the anonymous referee for their constructive comments, which helped to improve the quality of the
  manuscript. We also thank Richard Teague and Jeff Jennings for useful comments on using {\tt eddy} and {\tt
    frank}. This paper makes use of the following ALMA data: ADS/JAO.ALMA\#2017.1.01687.S,
  ADS/JAO.ALMA\#2017.1.01167.S, and ADS/JAO.ALMA\#2017.1.01419.S. ALMA is a partnership of ESO (representing
  its member states), NSF (USA) and NINS (Japan), together with NRC (Canada), MOST and ASIAA (Taiwan), and
  KASI (Republic of Korea), in cooperation with the Republic of Chile. The Joint ALMA Observatory is operated
  by ESO, AUI/NRAO and NAOJ. A.R. has been supported by the UK Science and Technology research Council (STFC)
  via the consolidated grant ST/S000623/1 and by the European Union’s Horizon 2020 research and innovation
  programme under the Marie Sklodowska-Curie grant agreement No. 823823 (RISE DUSTBUSTERS project).
  P.W. acknowledges support from FONDECYT grant 3220399.  S.P. acknowledges support from FONDECYT grant
  1191934. This work was funded by ANID -- Millennium Science Initiative Program -- Center Code NCN2021\_080.
  This project has received funding from the European Research Council (ERC) under the European Union Horizon
  2020 research and innovation program (grant agreement No. 101042275, project
  Stellar-MADE). A.A. acknowledges support through a Fellowship for National PhD students from ANID, grant
  number 21212094 and funding by ANID, Millennium Science Initiative, via the Núcleo Milenio de Formación
  Planetaria (NPF). P.R.M. thanks the Spanish MINECO for funding support from
  PID2019-106235GB-I00. C.C. acknowledges support by ANID BASAL project FB210003 and ANID, -- Millennium
  Science Initiative Program -- NCN19\_171. M.V. research was supported by an appointment to the NASA
  Postdoctoral Program at the NASA Jet Propulsion Laboratory, administered by Oak Ridge Associated
  Universities under contract with NASA.
\end{acknowledgements}

%-------------------------------------------------------------------
\bibliographystyle{aa}
\bibliography{biblio}

\begin{appendix}

\section{Continuum images at 0.89\,mm and 2.2\,mm}\label{appendix:continuum_bands7and4}

\begin{figure}
  \centering
  \includegraphics[width=\hsize]{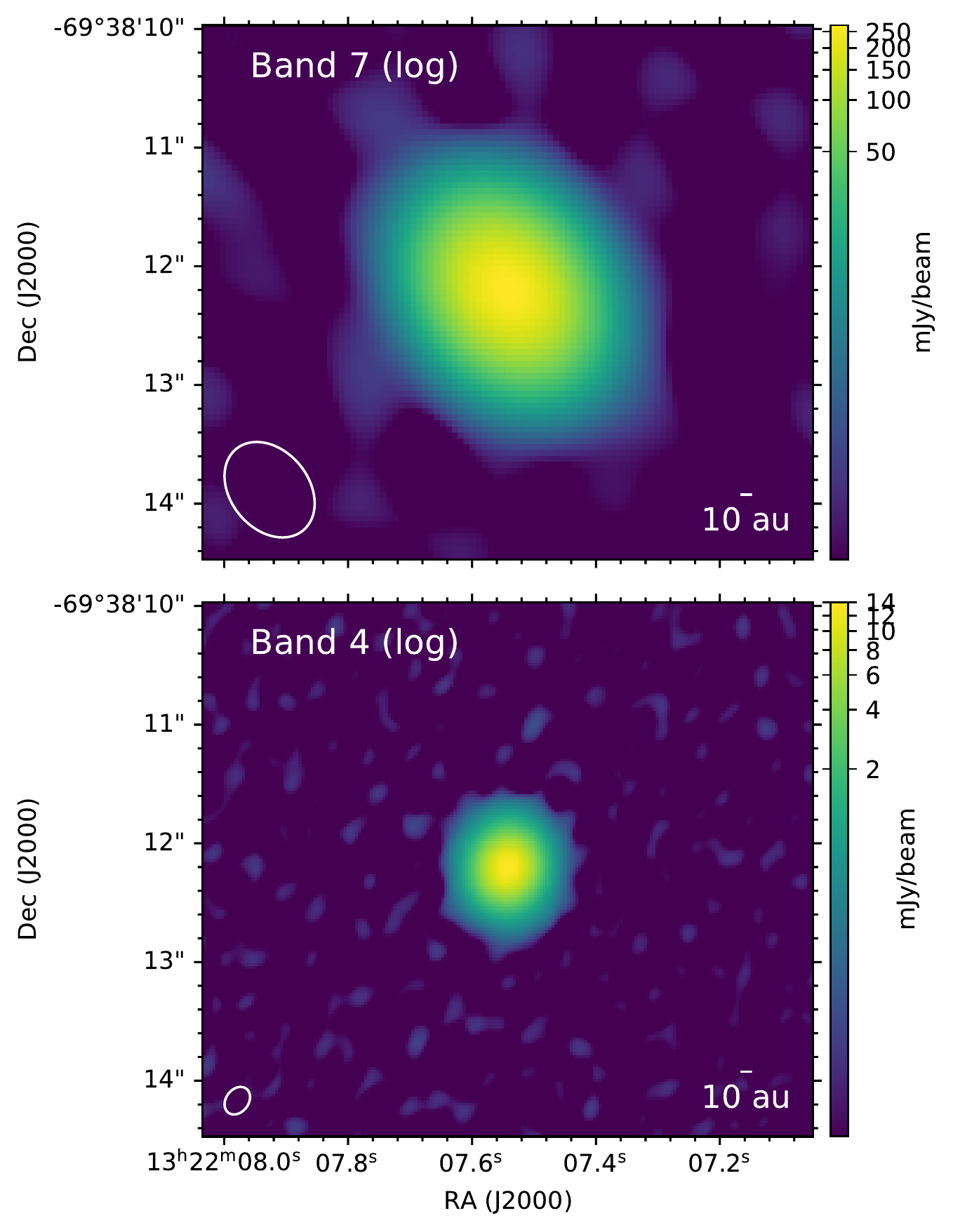}\hfill
  \caption{ALMA continuum images of MP~Mus at 0.89\,mm (top, unresolved) and 2.2\,mm (bottom, resolved) shown
    in logarithmic scales. The images were synthesized with a {\tt robust} value of -0.5. The corresponding
    beams (0.89\arcsec$ \times$0.66\arcsec and 0.25\arcsec$\times$0.19\arcsec at 0.89\,mm and 2.2\,mm,
    respectively) are shown at the bottom left corners as white ellipses.}\label{fig:continuum_bands7and4}
\end{figure}

The ALMA synthesized images at 0.89\,mm and 2.2\,mm are shown in Fig.~\ref{fig:continuum_bands7and4}.

\section{{\tt frank} visibility fit}

Figure~\ref{fig:frank_vis_fit} shows a comparison of the observed continuum visibilities at 1.3\,mm and the
corresponding fit derived with {\tt frank}.

\begin{figure}
  \centering
  \includegraphics[width=\hsize]{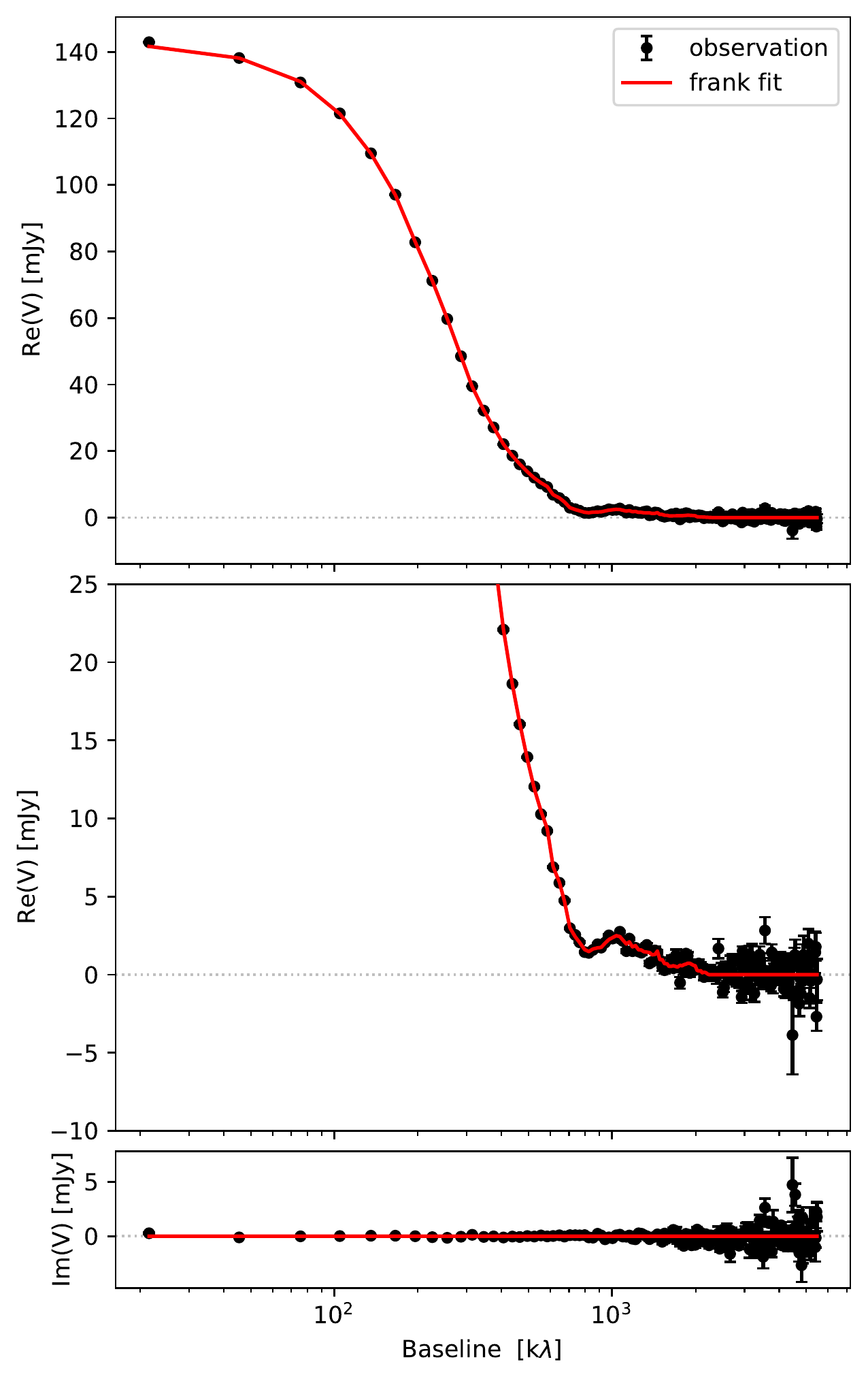}
  \caption{Comparison of the observed ALMA 1.3\,mm continuum visibilities of MP~Mus (black dots) and
      the results from the {\tt frank} fit (red line). The visibilities are shown in 30-k$\lambda$ bins.}
 \end{figure}\label{fig:frank_vis_fit}

  \section{Molecular gas emission line images}\label{appendix:lines}
\begin{figure*}
  \centering
  \includegraphics[width=\hsize]{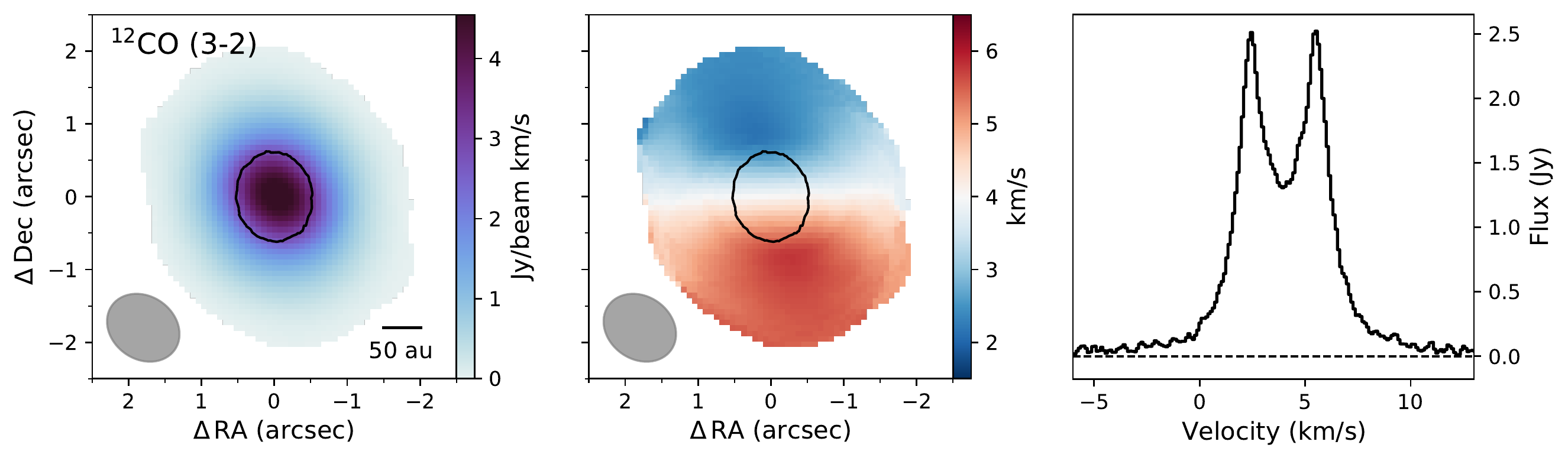}
  \includegraphics[width=\hsize]{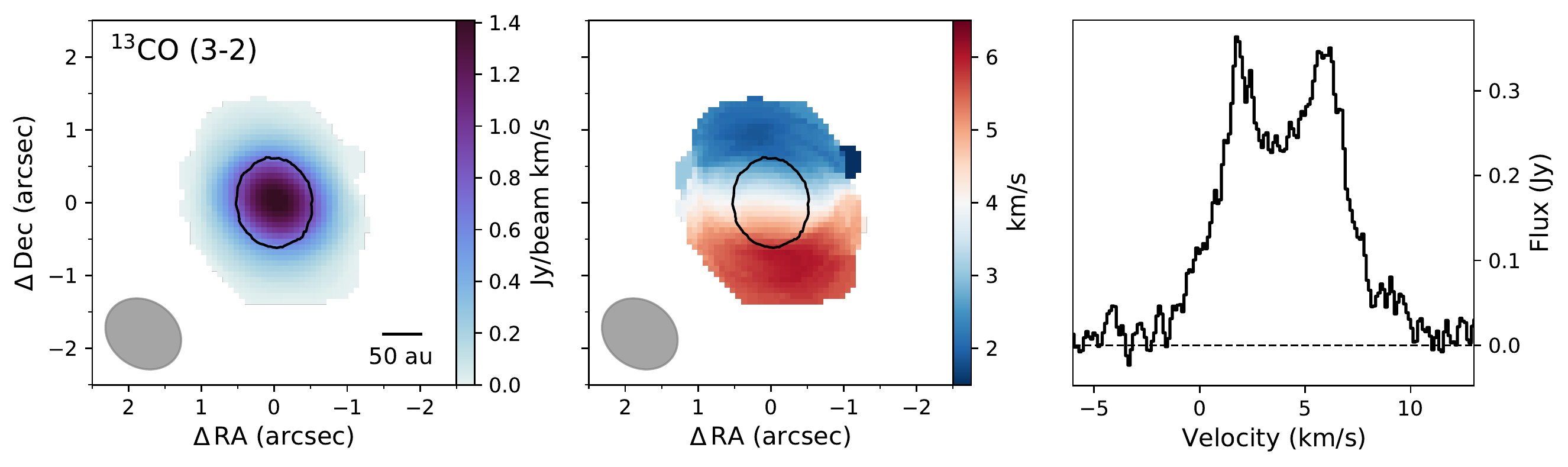}
  \includegraphics[width=\hsize]{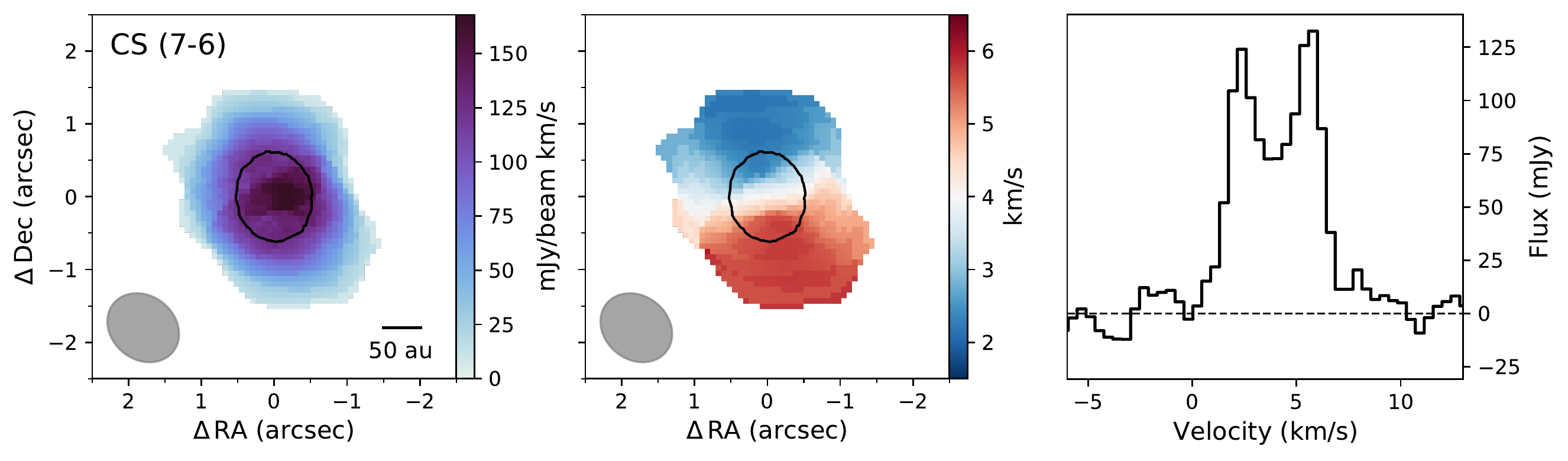}
  \includegraphics[width=\hsize]{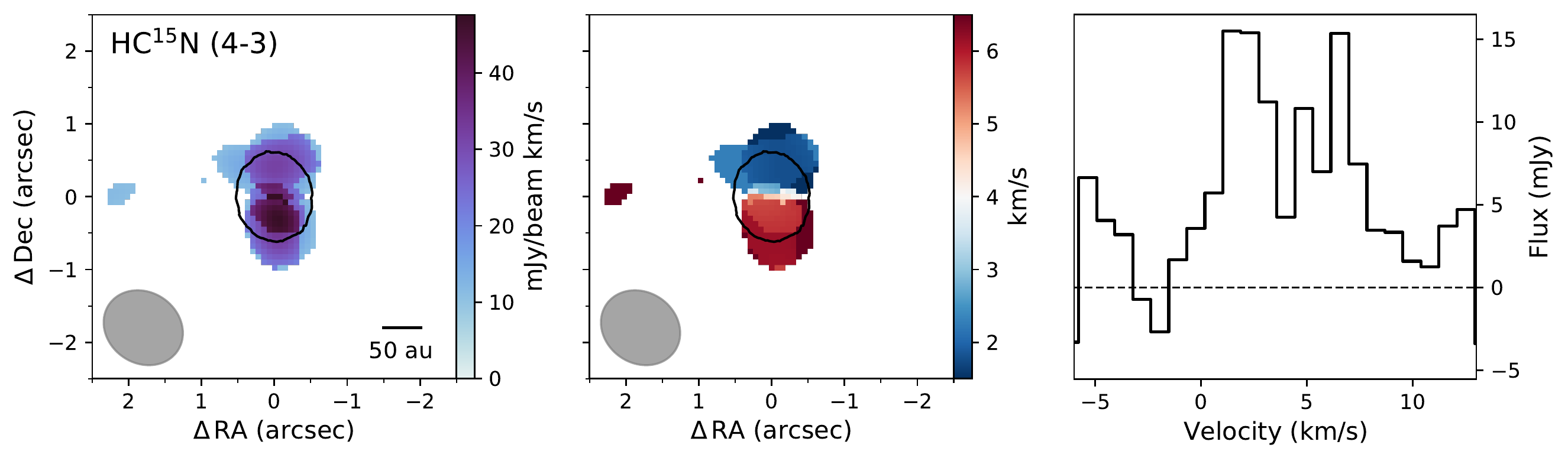}
  \caption{Moments and spectra of the molecular lines detected by ALMA in the MP Mus protoplanetary disk
    in Band 7. Figures and symbols are the same as in Fig.~\ref{fig:lines_B6}.}\label{fig:lines_B7}
\end{figure*}

\begin{figure*}
  \centering
  \includegraphics[width=\hsize]{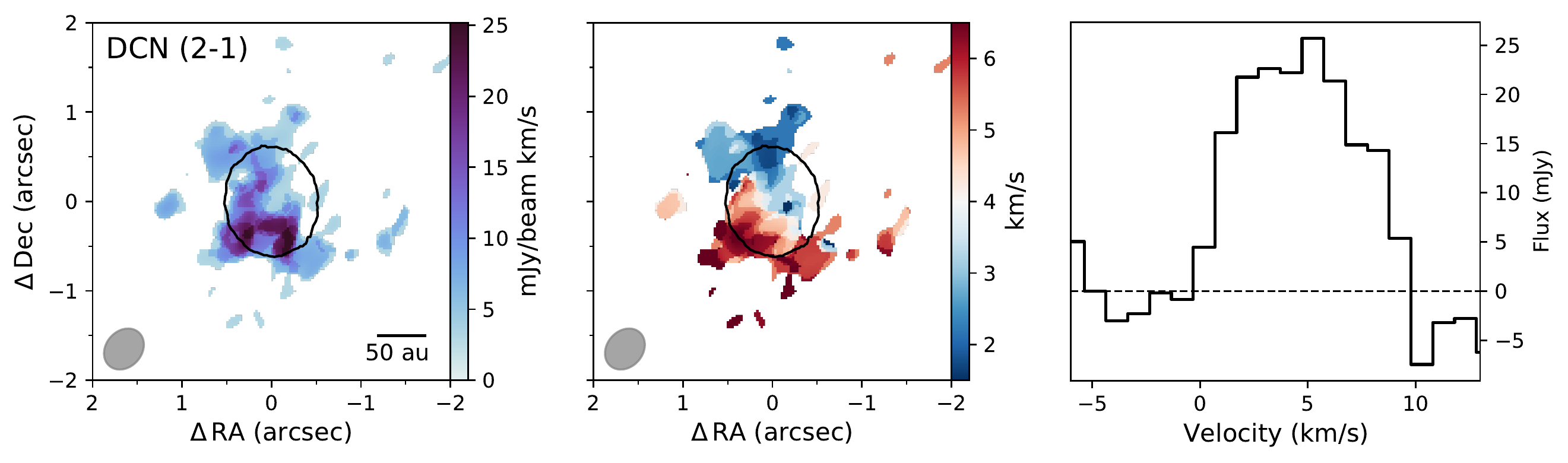}
  \includegraphics[width=\hsize]{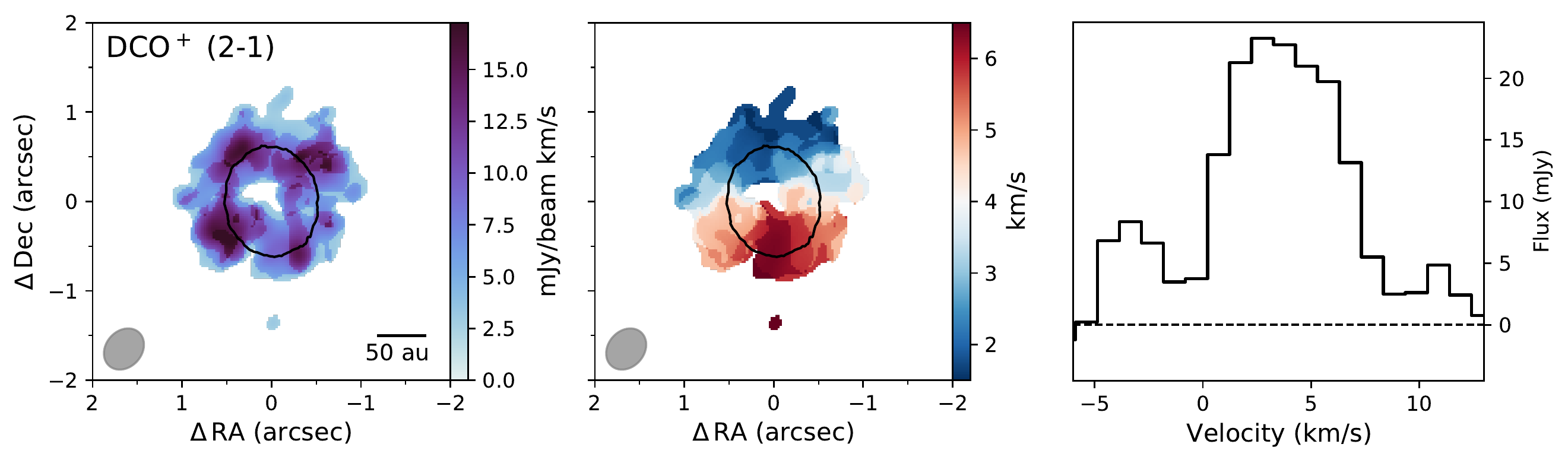}
  \includegraphics[width=\hsize]{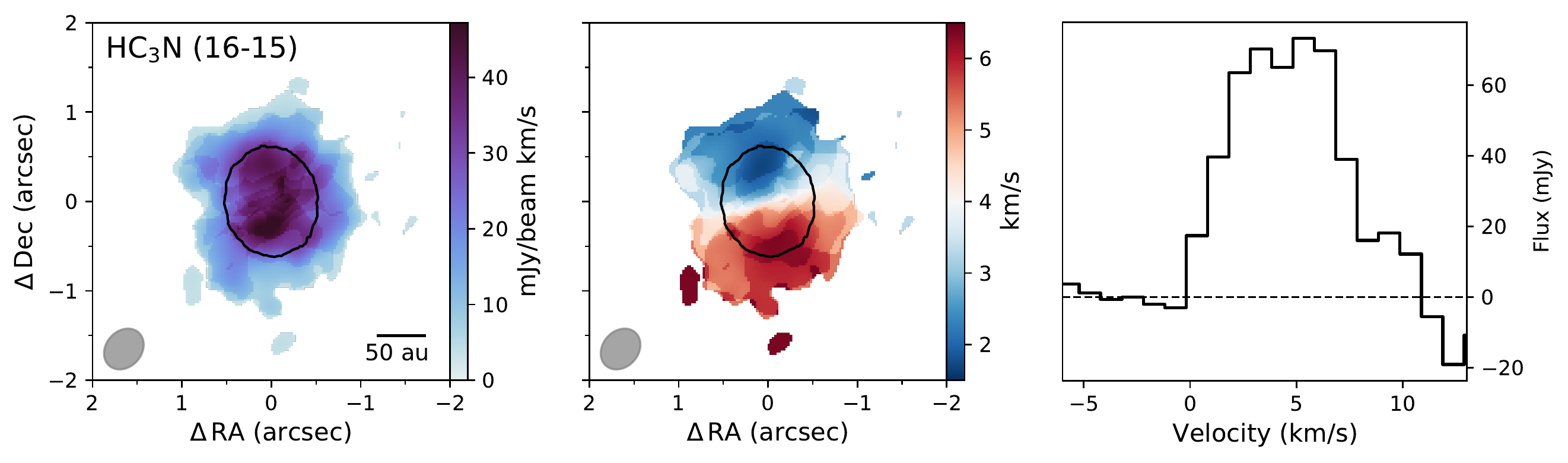}
  \caption{Moments and spectra of the molecular lines detected by ALMA in the MP Mus protoplanetary disk
    in Band 4. Symbols are the same as in Figs.~\ref{fig:lines_B6} and \ref{fig:lines_B7}.}\label{fig:lines_B4}
\end{figure*}

  Figures \ref{fig:lines_B7}, and \ref{fig:lines_B4}, show the zero-th and first moments
  of the molecular emission lines detected by ALMA in Bands 7 and 4, as well as their spectra.

\end{appendix}

\end{document}